\documentclass[useAMS,usenatbib]{mn2e}

\usepackage{graphicx, txfonts, natbib,color}
\usepackage{subfigure}
\usepackage{epsf}
\usepackage{amssymb}
\usepackage{epstopdf}
\usepackage{longtable,lscape}
\newcommand{\msun}{\mbox{M$_{\odot}$\,}}
\newcommand{\msol}{\mbox{M$_{\odot}$\,}}
\newcommand{\zsol}{\mbox{Z$_{\odot}$\,}}
\newcommand{\zsun}{\mbox{Z$_{\odot}$\,}}

\DeclareMathAlphabet{\mathsc}{OT1}{cmr}{m}{sc}
\def\testbx{bx}%
\DeclareRobustCommand{\ion}[2]{%
\relax\ifmmode
\ifx\testbx\f@series
{\mathbf{#1\,\mathsc{#2}}}\else
{\mathrm{#1\,\mathsc{#2}}}\fi
\else\textup{#1\,{\mdseries\textsc{#2}}}%
\fi}
\newcommand{\ha} {\mbox{H$\alpha$}\,}
\newcommand{\hb} {\mbox{H$\beta$}\,}
\newcommand{\hg} {\mbox{H$\gamma$}\,}
\newcommand{\hd} {\mbox{H$\delta$}\,}

\newcommand{\Caii} {[\ion{Ca}{ii}]\,}

\newcommand{\Feiii} {[\ion{Fe}{iii}]\,}
\newcommand{\Hii} {\ion{H}{ii}\,}
\newcommand{\Hei} {\ion{He}{i}\,}
\newcommand{\Heii} {\ion{He}{ii}\,}
\newcommand{\Ni} {[\ion{N}{i}]\,}
\newcommand{\Nii} {[\ion{N}{ii}]\,}
\newcommand{\Oi} {[\ion{O}{i}]\,}
\newcommand{\Oinir} {\ion{O}{i}\,}
\newcommand{\Oii} {[\ion{O}{ii}]\,}
\newcommand{\Oiii} {[\ion{O}{iii}]\,}
\newcommand{\Sii} {[\ion{S}{ii}]\,}

\newcommand{\Siii} {[\ion{S}{iii}]\,}

\newcommand{\Neiii} {[\ion{Ne}{iii}]\,}
\newcommand{\Ariii} {[\ion{Ar}{iii}]\,}
\newcommand{\Ariv} {[\ion{Ar}{iv}]\,}

\newcommand{\Mgi} {\ion{Mg}{i}\,}
\newcommand{\Mgii} {\ion{Mg}{ii}\,}

\def \aj {AJ}
\def \mnras {MNRAS}
\def \apj {ApJ}
\def \apjl {ApJL}
\def \aap {A\&A}
\def \nat {Nature}
\def \araa {ARAA}

\def \pasp {PASP}

\def \apjs {ApJS}
\def \aaps {A\&AS}

\title[The host galaxy and late-time evolution of the Super-Luminous Supernova PTF12dam]{The host galaxy and late-time evolution of the Super-Luminous Supernova PTF12dam}
\author[T.-W. Chen et al.]{T.-W.~Chen,$^1$\thanks{E-mail: tchen09@qub.ac.uk}
S. J.~Smartt$^1$, A. Jerkstrand$^1$, M. Nicholl$^{1}$, F. Bresolin$^{2}$, R. Kotak$^1$, \newauthor
J. Polshaw$^{1}$, A. Rest$^{3}$, R. Kudritzki$^2$, Z. Zheng$^4$, 
N. Elias-Rosa$^{5,6}$, K. Smith$^{1}$, C. Inserra$^{1}$, \newauthor
D. Wright$^{1}$, E. Kankare$^{1,7}$, T. Kangas$^8$,
M. Fraser$^{9}$
 \\
$^1$Astrophysics Research Centre, School of Maths and Physics, Queen's University Belfast, Belfast BT7 1NN, UK\\
$^2$Institute for Astronomy, 2680 Woodlawn Drive, Honolulu, HI 96822, USA\\
$^3$Space Telescope Science Institute, 3700 San Martin Dr., Baltimore, MD 21218, USA\\
$^4$National Astronomical Observatories, Chinese Academy of Sciences, Beijing 100012, China\\
$^5$INAF-Osservatorio Astronomico di Padova, Vicolo dell'Osservatorio 5, 35122 Padova, Italy\\
$^6$Institut de Ci{`}encies de l’Espai (CSIC-IEEC), Campus UAB, Torre C5, 2a planta, 08193 Barcelona, Spain\\
$^7$Finnish Centre for Astronomy with ESO (FINCA), University of Turku, V\"ais\"al\"antie 20, FI-21500 Piikki\"o, Finland\\
$^8$Tuorla Observatory, Department of Physics and Astronomy, University of Turku, V\"ais\"al\"antie 20, FI-21500 Piikki\"o, Finland\\
$^{9}$Institute of Astronomy, University of Cambridge, Madingley Road, Cambridge, CB3 0HA, UK\\
}

\begin{document}

\date{17 June 2015}

\pagerange{\pageref{firstpage}--\pageref{lastpage}} \pubyear{2015}

\maketitle

\label{firstpage}

\begin{abstract}
Super-luminous supernovae (SLSNe) of type Ic have a tendency to occur in faint host galaxies which are likely to have low mass and low metallicity. PTF12dam is one of the closest and best studied super-luminous explosions that has a broad and slowly fading lightcurve similar to SN~2007bi. 
Here we present new photometry and spectroscopy for PTF12dam from 200-500 days (rest-frame) after peak and a detailed analysis of the host galaxy (SDSS J142446.21+461348.6 at $z = 0.107$). Using deep templates and image subtraction we show that the lightcurve  can be fit with a magnetar model if escape of high-energy gamma rays is taken into account. 
The full bolometric lightcurve from $-53$ to +399 days (with respect to peak) cannot be fit satisfactorily with the pair-instability models. An alternative model of interaction with a dense CSM produces a good fit to the data although this requires a very large mass ($\sim 13$\,\msun) of hydrogen free CSM. The host galaxy is a compact dwarf (physical size $\sim 1.9$ kpc) and with $M_{g} = -19.33 \pm 0.10$, it is the brightest nearby SLSN Ic host discovered so far. The host is a low mass system ($2.8 \times 10^{8}$ \msun) with a star-formation rate (5.0 \msun\,yr$^{-1}$), which implies a very high specific star-formation rate (17.9 Gyr$^{-1}$).
The remarkably strong nebular emission provide detections of the \Oiii $\lambda4363$ and \Oii $\lambda\lambda7320,7330$ auroral lines and an accurate oxygen abundance of $12 + \log{\rm (O/H)} = 8.05 \pm 0.09$. 
We show here that they are at the extreme end of the metallicity distribution of dwarf galaxies and propose that low metallicity is a requirement to produce these rare and peculiar supernovae. 
\end{abstract}


\section{Introduction}
\label{sec:introduction}

New types of violent explosion labelled super-luminous supernovae (SLSNe) have been discovered by the current generation of wide-field optical surveys. These SLSNe are 10 to 100 times brighter than normal core-collapse SNe (CCSNe) and reach absolute magnitudes above {\it M} $< -21$ \citep[for a review, see][]{2012Sci...337..927G}. They have attracted considerable attention due to their potential utility as cosmological standard candles \citep[see][for the first result on a standardisation]{2014arXiv1409.4429I}. However, the mechanism which powers such luminosities is still not well established. Some of these SLSNe show signs of strong interaction between high velocity ejecta from a SN explosion and pre-existing dense circumstellar material. In such cases prominent lines of hydrogen are seen with multiple velocity components
\citep[e.g.][]{2007ApJ...671L..17S,2008ApJ...686..467S,2014MNRAS.441..289B} and the source of the high and long lasting luminosity is almost certainly reprocessing of the kinetic energy of the 
ejecta during collisions with slower moving shells. Another group of SLSNe have been called SLSNe I or SLSNe Ic (we will use the latter term in this paper), 
since they do not generally exhibit hydrogen or helium lines in their optical spectra, and show no spectral signatures of interaction \citep[e.g.][]{2010ApJ...724L..16P,2011Natur.474..487Q,2011ApJ...743..114C,2013ApJ...770..128I}. The first of these to be discovered was SN~2005ap
\citep{2007ApJ...668L..99Q} which, when linked to the early Palomar Transient Factory (PTF) discoveries by \cite{2011Natur.474..487Q}, defined the class of these remarkable explosions. 
Some physical causes of the luminosity have been proposed. One is a model of spinning down magnetic neutron stars
\citep{2010ApJ...717..245K,2010ApJ...719L.204W} and quantitative fits to some of these show good agreement \citep{2011ApJ...743..114C,2013ApJ...770..128I}. However, there are alternatives and the interaction model for these hydrogen-poor events is still plausible even though 
there are no obvious spectral signatures \citep{2011ApJ...729L...6C}. 

A few of these SLSNe Ic have lightcurves that decay quite slowly and
the prototype of the class is SN~2007bi \citep{2009Natur.462..624G, 2010A&A...512A..70Y}. 
\citet{2009Natur.462..624G} proposed that SN~2007bi was the result of a pair-instability supernova (PISN) explosion and the slowly fading lightcurve is driven by the decay of a large mass of $^{56}$Ni in the ejecta. 
PTF10mnm is also a probable PISN candidate, and its lightcurve shape and peak luminosity were fitted well with a massive PISN model from \citet{2014A&A...565A..70K}. 
However, other works e.g. \citet{2013Natur.502..346N} and \cite{2014MNRAS.437..656M} have recently shown that two SLSNe Ic (PTF12dam and PS1-11ap) have very similar 
observational characteristics to SN~2007bi, presenting better data before and around peak. They argue that the detailed lightcurve fits to both the rise time and decay time cannot match the models of pair-instability and that 
$^{56}$Ni is not the cause of the extreme luminosity. \citet{2013Natur.502..346N} 
and \cite{2014MNRAS.437..656M} argued that the observational data of both of these SLSNe Ic could be better explained by 
magnetar powering. The key point supporting magnetar engines rather than pair-instability explosions, is that the pre-maxima data show a steep rise that is not consistent with pair-instability models, but are well fit by magnetar models with 10-16 \msun ejecta. This suggests that SN~2007bi may also not actually be a pair-instability explosion. 
This view was also expressed by the quantitative modelling of \citet{2010ApJ...717..245K} and \citet{2012MNRAS.426L..76D} who both suggested a magnetar as an alternative power source
to a large mass of radioactive $^{56}$Ni. Another mechanism for boosting the luminosity of normal supernovae is accretion onto compact remnants as recently suggested by \citet{2013ApJ...772...30D}. 

All of these alternative models (magnetars, CSM interaction, radioactivity, accretion) to interpret SLSNe Ic have some success in reproducing the data sets currently available. Due to the relatively low volumetric rate of SLSNe Ic \citep[e.g.][]{2013MNRAS.431..912Q, 2014MNRAS.437..656M}, the closest events have been found at redshifts between 0.1 and 0.3, and the observational data sets typically cover the evolution of the SNe in the ultraviolet (UV), optical and near-infrared (NIR) from around 30 days before peak to 100-200 days afterwards (rest-frame). Distinguishing between models would benefit from wider wavelength coverage \citep[e.g. in the X-rays as shown by][]{2014MNRAS.437..703M} or observing SNe very late in their evolution. 
\citet{2009Natur.462..624G} showed how the nebular spectrum of SN~2007bi (at +470d) could be used to support a pair-instability
interpretation. \cite{2013Natur.502..346N} argued that while this was a consistent argument, 
one could also model the observed line flux with a much lower ejecta mass than that 
required for pair-instability. \citet{2013ApJ...763L..28C} illustrated the use of 
deep imaging at late times to set limits on the amount of $^{56}$Ni explosively produced 
in SLSNe. An upper limit 
 of $^{56}$Ni mass $\leq$ 0.4 \msol was found for SN~2010gx \citep{2010ApJ...724L..16P,2011Natur.474..487Q} at 240-560d after explosion. 
The studies of \citet{2013ApJ...770..128I} 
and \cite{2014MNRAS.444.2096N} further attempted to recover the fading stages of SLSNe Ic and
fit the luminosity with the magnetar model. 

SLSNe Ic appear to have a preference for occurring in faint, dwarf galaxies \citep{2011ApJ...727...15N}, and they also show a trend towards low metallicity environment \citep{2011ApJ...730...34S, 2013ApJ...763L..28C, 2013ApJ...771...97L}. The most reliable method to quantitatively determine oxygen abundance 
without calibration uncertainties is the ``direct method'' which requires the electron temperature $T_{\rm e}$ to be estimated from the
auroral \Oiii $\lambda4363$ line. The first detailed study of a SLSN Ic host (SN~2010gx)
using the detection of this line \citep{2013ApJ...763L..28C} suggested that low metallicity plays a critical role in producing SLSNe Ic. They found the host of SN~2010gx to have the lowest metallicity of any SN to date; the oxygen abundance of $12 + \log{\rm (O/H)} = 7.4 \pm 0.1$ (employing the direct $T_{\rm e}$ method) is equivalent to 0.05 \zsol for a solar value of $12 + \log{\rm (O/H)} = 8.69 \pm 0.05$ \citep{2009ARA&A..47..481A}. \citet{2013ApJ...771...97L} found the metallicity of the host of PS1-10bzj to be $12 + \log{\rm (O/H)} = 7.8 \pm 0.2$ (also using the $T_{\rm e}$ method with a detection of the weak [O\,{\sc iii}] 4364 line). This host has the highest specific star-formation rate (sSFR) $\sim 100$ Gyr$^{-1}$ among all SLSN Ic hosts so far. The stellar mass of the hosts of SLSNe Ic tend to fall below $10^{9}$ \msun, hence their sSFR are quite high even for dwarf galaxies in the local Universe \citep{2013ApJ...763L..28C, 2014ApJ...787..138L}. 

Low metallicity may be a major constraint on the progenitor channel of SLSNe Ic, if it is a common and exclusive feature among the dwarf galaxy hosts. 
However the auroral \Oiii$\lambda4363$ line is often weak and at redshifts between 0.1 and 0.5 is difficult to detect. Spectroscopy with 10m-class telescopes is needed and even then the 
emission line strengths of the hosts vary considerably such that the detection of this line is not 
always possible. The 
more commonly used metallicity diagnostic is the $R_{23}$ strong line method, based on the ratio of \Oiii$\lambda\lambda4959,5007$/\hb. The majority of estimates of metallicity for the sites of SLSNe Ic to date \citep{2014ApJ...787..138L, 2010A&A...512A..70Y, 2014MNRAS.437..656M} are from measuring the strong nebular line \hb, and the forbidden \Oiii/\Oii lines and applying a calibration such as \citet{1991ApJ...380..140M} (hereafter M91). However it has been known for some time that the calibration of the strong line methods vary by nearly an order of magnitude. Typically there are systematic offsets of 0.2-0.4 dex (and sometimes up to 0.6 dex) between the $R_{23}$ calibration and abundances determined on an electron temperature scale \citep{2007A&A...473..411L, 2011ApJ...729...56B}. An additional complication for the $R_{23}$ calibration is that there are two branches
of the calibration curve and one needs to know which one to apply. 
Hence it is desirable, where possible, to measure the strength of the weak \Oiii$\lambda4363$ line. The most extensive study to date of 
SLSN Ic hosts is by \citet{2014ApJ...787..138L} who found that 31 hosts are in general low mass ($\sim 10^{8}$ \msun), low-luminosity and low-metallicity ($\sim 0.45$ \zsol) based on $R_{23}$ method, and have a high median sSFR ($\sim 2$ Gyr$^{-1}$). They suggested the SLSN Ic host population is similar to long gamma-ray burst (LGRB) hosts.
Recently, \citet{2015MNRAS.449..917L} suggested that, while the host galaxies are similar for SLSNe Ic and LGRBs, SLSN hosts are in fact more extreme. They found that half of their sample of SLSN Ic hosts are extreme emission line galaxies (having emission lines with EW $>$ 100\AA), and claimed that the progenitors of SLSNe Ic are the first generation of stars to explode in a starburst - even younger than GRBs. In contrast, \citet{2015ApJ...804...90L} found that the locations of SLSNe within their hosts do trace the UV light, and thus are correlated with recent star formation (massive progenitors), but they appear to have less of a preference for occurring in the brightest regions than do LGRBs (e.g. \citealp{2014ApJ...789...23K}). Hence they suggested that SLSN Ic progenitors are older/less massive stars than those of LGRBs.

As PTF12dam is one of the closest SLSNe Ic ever found at redshift $z = 0.107$ and this provides an opportunity to study a host galaxy in detail. This galaxy has 
exceptionally strong emission lines and with auroral lines of three atomic species available
one can determine abundances more precisely than done before. 
In this paper we present imaging and spectroscopy data at very late epochs in the SN evolution; the multi-wavelength lightcurves sample out to 400 days (rest-frame) after peak and nebular spectra are presented at $\sim$ 500 days. A major complication with this observational experiment is that SLSNe Ic are usually found in faint, but compact, host galaxies and hence distinguishing between SN flux and galaxy flux requires careful long term monitoring. The host galaxies and explosion environments are themselves of interest and a study of the late time evolution is necessarily aligned with quantifying the host galaxy contribution. The requirement to separate host galaxy flux from SN flux necessitates a careful joint analysis of these two flux. 

Since PTF12dam has a slowly fading lightcurve like SN~2007bi and PS1-11ap, we also compared these three SN host galaxies. 
The paper is organised as follows: 
in section 2, we detail the photometric follow-up and spectroscopic observations of PTF12dam and its host galaxy. We also present new photometric data of the host of SN~2007bi. Section 3 discusses the various host galaxy properties, the methods to determinate the host stellar mass, metallicity and star-formation rate, etc. Section 4 investigates the bolometric lightcurve of PTF12dam from $-52$ to +399 days. 
Our discussion on lightcurve modelling is presented in section 5, and on the host in section 6. Finally, we conclude in section 7.

\begin{figure*}
       \centering
       \begin{subfigure}
           \centering
           \includegraphics[angle=0,width=100pt]{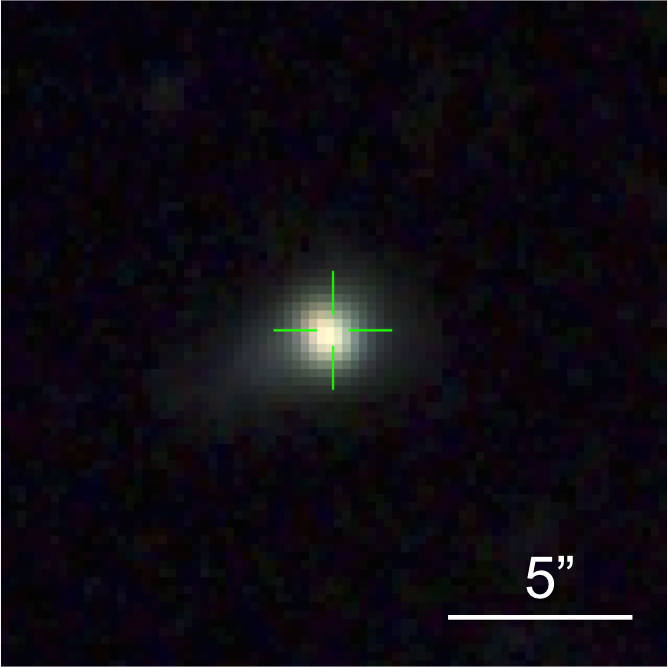}
       \end{subfigure}
       \begin{subfigure}
           \centering
           \includegraphics[angle=0,width=100pt]{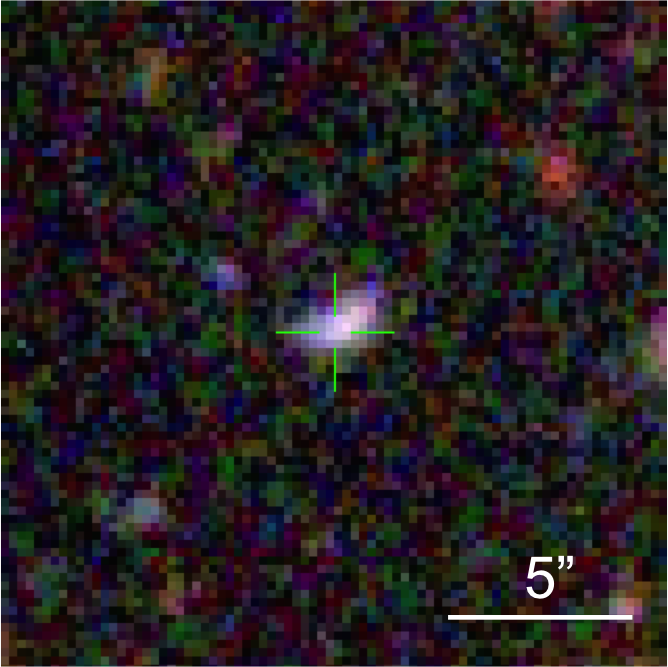}
       \end{subfigure}
       \begin{subfigure}
           \centering
           \includegraphics[angle=0,width=100pt]{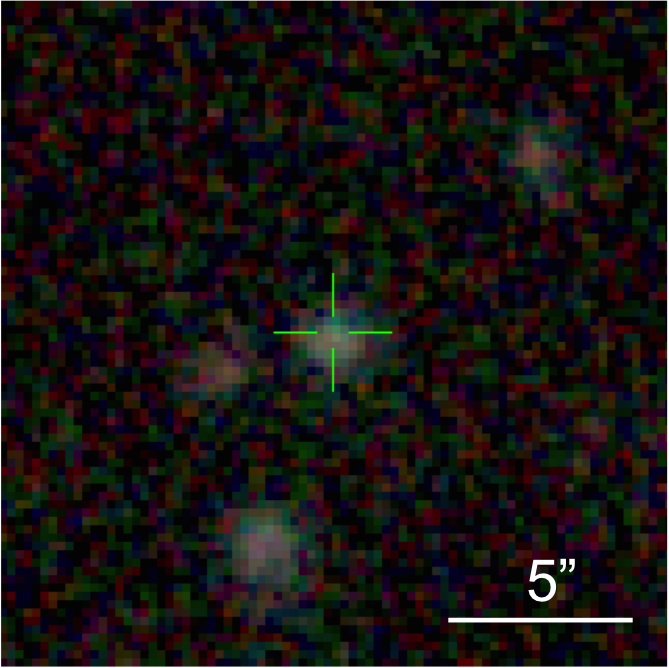}
       \end{subfigure}
       \caption{RGB-colour combined host galaxy images. The SN position is marked with green crosshairs and was determined by superimposing the SN image at peak brightness to the host galaxy frame \citep{2010A&A...512A..70Y, 2013Natur.502..346N, 2014MNRAS.437..656M}. The images are 20\arcsec$\times$20\arcsec with North pointing up and East to the left. 
{\it Left}: The host of PTF12dam combined with {\it gri}-band images. {\it Central}: The host of SN~2007bi combined with 
{\it gri}-band images. {\it Right}: The host of PS1-11ap combined with Pan-STARRS1 {\it r$_{\rm P1}i_{\rm P1}z_{\rm P1}$} filter images.}
\label{fig:host_colour}
\end{figure*}

\section[]{Observational data and analysis}
\label{sec:observational_data}

\subsection{Host galaxy photometry}
\label{sec:host_galaxy_photometry}

The host of PTF12dam is a clear source detected in the Sloan Digital Sky Survey (SDSS) images (SDSS J142446.21+461348.6) taken on 2003 February 11, about 10 years before the explosion of PTF12dam. While we have gathered deeper and higher resolution images of PTF12dam and its host at $\sim$ 2 years after the SN explosion, we used the SDSS magnitudes as the most appropriate for the host galaxy since they unambiguously do not contain any supernova light. We adopted the apparent Petrosian AB magnitudes {\it u} = $19.79\pm0.05$, {\it g} = $19.36\pm0.02$, {\it r} = $19.20\pm0.02$, {\it i} = $18.83\pm0.03$ and {\it z}=$19.32\pm0.14$ from the SDSS version 9 data release catalog (DR9) \citep{2012ApJS..203...21A} for the PTF12dam host galaxy. The host is also detected in the Galaxy Evolution Explorer (GALEX) images, having the UV magnitudes of {\it FUV} = $20.13\pm0.19$ ($\sim$ 1528 \AA) and {\it NUV} = $20.13\pm0.14$ ($\sim$ 2271 \AA) in the AB system (GR6 catalog\footnote{http://galex.stsci.edu/GR6/}), which corresponds to the object GALEX J142446.2+461348.

The host galaxy of PTF12dam is not detected in the Two Micron All Sky Survey (2MASS) survey, its brightness is below the sensitivity limits of {\it J} = 17.0, {\it H} =16.0, {\it K} = 15.5 \citep[see supplementary information in ][]{2013Natur.502..346N} and also is not detected in the available Wide-field Infrared Survey Explorer (WISE) \citep[limit 17.23 mag at 3.4 $\mu$m;][]{2010AJ....140.1868W} or {\it Spitzer Space Telescope} images. Hence we lack any NIR pre-supernova constraints on the host. 
However, we obtained deep {\it JHK$_{s}$} images with the 2.6-m Nordic Optical Telescope (NOT) on 2014 February 13, which was +554d past peak (rest-frame). 
The NIR camera on NOT, NOTCam, is a $1024\times 1024$ pixel array providing a field of view of $4\times 4$ arcmin with 0\farcs23 pixels.
Those images were reduced by the external {\sc iraf/notcam} package which applies flatfielding and sky background subtractions. 
Aperture photometry within the {\sc iraf/daophot} package was carried out using an aperture of $\sim$ 3\arcsec, which encompassed the whole galaxy. We used the same aperture size to measure the flux of two 2MASS reference stars
in the field to set the zeropoint in each band. 
To ensure better relative photometry amongst the other epoch measurements, we added 
three other stars which were detected in the NOT images as local secondary standards. 
This led to derived NIR magnitudes of the host of PTF12dam of 
{\it J}$=18.37\pm0.15$, {\it H}$=17.82\pm0.13$ and {\it K$_{s}$}=$17.16\pm0.09$ in the Vega system. 
We assumed that the detected flux in the NOT NIR images is exclusively host galaxy, with 
negligible contribution from the supernova flux. As discussed below, the
measured {\it J}-band flux of PTF12dam (with image subtraction) is almost 2 magnitudes fainter than the 
host at +342d after peak and at a decline rate of roughly 1 magnitude per 100 days, we 
would expect it to be $\sim$ 4 magnitudes fainter than the measured host flux at this epoch of +553d (and even fainter in {\it H} and {\it K$_{\rm s}$}, see Table\,\ref{tab:phot})

Additionally, as PTF12dam is similar to SN~2007bi \citep{2013Natur.502..346N}, we have pursued further investigation of the host of SN~2007bi (SDSS J131920.14+085543.7). 
The SDSS detection published in \cite{2010A&A...512A..70Y}
is marginal in {\it g} and {\it r} and there are no flux detections in the other bands. \cite{2010A&A...512A..70Y} published estimates of the 
late time photometry of SN~2007bi and its host galaxy from the ESO 8-m Very Large Telescope (VLT) and the 2-m Liverpool Telescope (LT) \citep{2004SPIE.5489..679S}. However disentangling the host from the SN flux was still not unambiguous.
Therefore, we took deep images on the 4.2m William Herschel Telescope (WHT) with the auxiliary-port camera (ACAM) instrument on 2012 May 25. This camera provides imaging over 8 arcmin (0\farcs25 per pixel) with a low fringing deep depleted 2k $\times$ 4k EEV CCD. Images were taken through the {\it griz} filters on ACAM, (specifically filters : 701 SlnG, 702 SlnR, 703 SlnI, 704 SlnZ). These were reduced in standard fashion by de-biassing and flat-fielding with twilight sky frames. Zeropoints were determined with 10 SDSS reference stars in the field, and aperture photometry was carried out within {\sc iraf/daophot} using an aperture of $\sim$ 2.3\arcsec to cover the whole galaxy. 
We estimated apparent AB magnitudes {\it g} = $22.84\pm0.10$, {\it r} = $22.36\pm0.08$, {\it i} = $22.23\pm0.09$ and {\it z} = $21.98\pm0.13$ for the host of SN~2007bi. Colour images of each are shown in Fig.\,\ref{fig:host_colour}.

\subsection{Host galaxy spectroscopy}
\label{sec:host_galaxy_spectroscopy}

A large number of spectra of PTF12dam have been presented in 
\cite{2013Natur.502..346N} showing the very strong emission line nature of SDSS J142446.21+461348.6. We took additional spectra of PTF12dam since that paper was submitted and the main data we will focus on here for host galaxy analysis was taken on 2013 March 30. At this date, the SN was +265d after peak (rest-frame). The spectrum was taken with the WHT + Intermediate dispersion Spectrograph and Imaging System (ISIS), listed in Table\,\ref{tab:spectroscopy}. The dichroic was removed and the spectra were taken in the red and blue arms separately, since the \hb line at $z = 0.107$ falls almost right on the dichroic cross-over point, leading to uncertain flux calibration. 
In the red arm, the R158R grating was used (dispersion = 1.8 \AA/pixel) with an order-blocking filter GG495 and central wavelength of 7500 \AA, giving coverage 5075-9170 \AA.
We used a 1\farcs0 slit to obtain three spectra of 1800 sec exposures each. The object position was shifted by $\sim$ 10 pixels along the slit for each individual frame, allowing us to remove the sky emission lines by two-dimensional image subtraction. The grating R300B (dispersion = 0.86 \AA/pixel) was used for the blue arm with a central wavelength 4500 \AA\,to cover 3200-5985 \AA, 
and again a 1\farcs0 slit was used and three 1800 sec exposures taken. All observations were taken at the parallactic angle. To ensure that the spectra could be corrected for any slit losses, we took additional spectra with a 10\farcs0 slit in both red and blue arms for covering the total flux of the galaxy (FWHM $\sim$ 1.5\arcsec). 
The exposure times of these spectra were 600 sec each. 
In order to calibrate the flux carefully, we took two spectrophotometric standards, HZ44 and Feige 66, before and after the host observations, again with both the 1\farcs0 and 10\farcs0 slits. The wavelength calibrations were achieved using daytime CuNe+CuAr arcs. Detrending of the data, such as bias subtraction, and lamp flat-fielding was achieved using standard techniques within {\sc iraf}.

\begin{figure}
\includegraphics[angle=0,width=\linewidth]{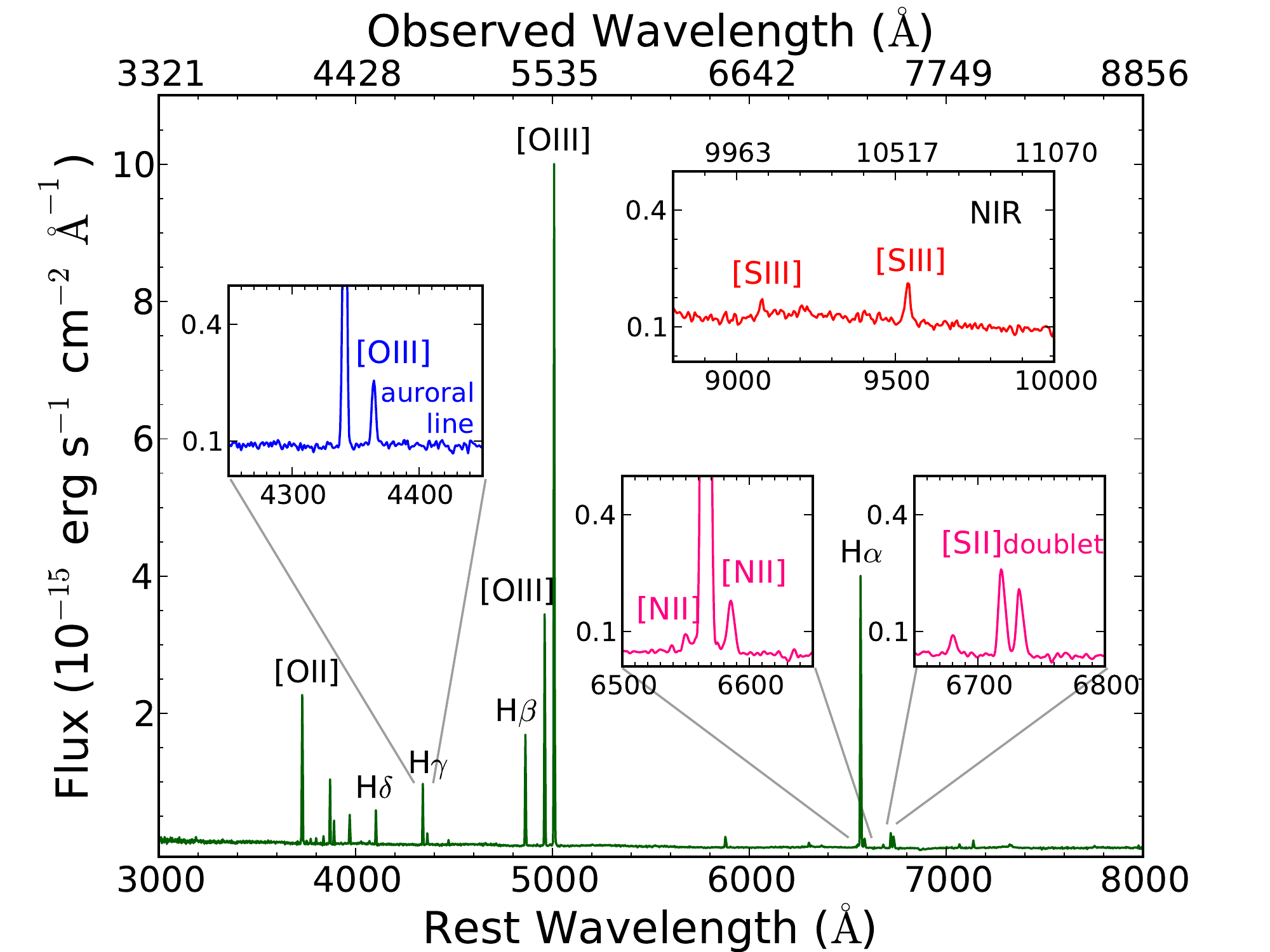}
\caption{Spectrum of the host galaxy of PTF12dam. The \Oiii $\lambda4363$ auroral line and the region near \ha are amplified. The NIR spectrum taken from \citet{2013Natur.502..346N} is shown in the upper-right panel.}
\label{fig:12dam_host_spec}
\end{figure}

The combined blue and red spectra provide spectral coverage between 3200 and 9170 \AA\,(rest-frame 2891-8284 \AA) and the overlap region was used to ensure a uniform flux calibration, employing the three strong lines \Oiii $\lambda$5007, \Oiii $\lambda$4959, and \hb. We measured the line flux in the red and blue arm spectra and applied a linear scaling (a scaling of 1.39 for red arm; 1.24 for blue arm) to bring both the 1\farcs0 slit spectra into agreement with the 10\farcs0 slit spectra. The line flux measured in the separate blue and red arms of these three lines agreed to within $1\%$ after this rescaling. Fig.\,\ref{fig:12dam_host_spec} shows the final combined spectrum of the host galaxy. 
Although the spectrum contains flux from PTF12dam, the strength of the emission lines are secure
given their strength and the fact that we subtract off the continuum. 
A wealth of emission lines were detected in the host. These include the auroral \Oiii $\lambda4363$ line, which is essential for the most reliable metallicity calibration by estimating the electron temperature of the ionised gas from the flux ratio with \Oiii $\lambda\lambda4959,5007$. The 
auroral \Oii doublet $\lambda \lambda$7320,7330 was also a strong detection. 
The abundance analysis of these spectra is discussed in section\,\ref{sec:metallicity}.

\begin{table*}
\begin{center}
\caption{Log of spectroscopic observations of PTF12dam and its host galaxy used in this paper. The WHT spectrum at $-13.6$d, and the TNG spectrum,  are from \citet{2013Natur.502..346N}. The resolution was estimated from the FWHM of sky lines.}
\label{tab:spectroscopy}
\begin{tabular}[t]{lllllllll}
\hline
Date & MJD & Phase (day) & Telescope & Instrument + Grism/Grating & Exposure (sec) & Slit (\arcsec) & Resolution (\AA) & Range (\AA) \\
\hline
2012 May 25 & 56072.91 & $-13.6$ & WHT & ISIS + R300B & 900 & 0.7 & 3 & 3000-5400 \\
2012 Jul 09 & 56117.99 & 27.1 & TNG  & NICS + IJ & $600 \times 6$ & 1.5 & 35 & 8700-13500 \\ 
2013 Mar 30  & 56382.07  & 265.6 & WHT & ISIS + R300B + R158R & $1800 \times 3 $ & 1 & 4, 6 & 3200-9170\\ 
2013 Mar 30  & 56382.09  & 265.6 & WHT & ISIS + R300B + R158R & 600  & 10 & - & 3200-9170\\ 
2013 Dec 25 & 56651.20    & 508.8 & GTC & OSIRIS + R500R        & $2400 \times 3$ & 1 & 16 & 4900-9200 \\
\hline
\end{tabular}
\end{center}
\end{table*}

Line flux measurements were made after fitting a polynomial function to the continuum and subtracting off this contribution. This continuum flux is composed of the host galaxy continuum from its constituent stellar population and the supernova flux from PTF12dam which is still visible. 
We fitted gaussian line profiles using the QUB custom built {\sc procspec} environment within {\sc IDL}. The full width at half maximum (FWHM) was allowed to vary for the gaussian fits for the strong lines (defined as their  equivalent width (EW) $\geq$ 10 \AA) and the unresolved blends; while we fixed the FWHM of single, weak lines (those with EW $<$ 10 \AA). 
The resolutions for the R300B and R158R gratings, for a 1\farcs0 slit are similar (R $\simeq$ 1200 and 1000 respectively) and we measured the arclines to have FWHM = 3.7 \AA\ in the blue and 
FWHM = 6.6 \AA\ in the red. 
The single transition lines are marginally resolved. 
For the weak lines (EW $< 10$ \AA), we estimated the average velocity
from the three strongest lines (\Oiii $\lambda$5007, \Oiii $\lambda$4959, and \hb) and then determined the appropriate FWHM to fix 
at the observed central wavelength of each. 
In addition, we fixed the widths of both \Nii $\lambda\lambda6548,6583$ lines in order to more reliably measure
their flux and to discriminate from the wings of the nearby strong \ha line (even though the 6583 \AA\ line was stronger than 10 \AA). 
We did not fix the FWHM of the weak but clearly blended lines such as the \Feiii $\lambda4658$ doublet and we also accepted the best-fitting FWHM for the strong \Oii $\lambda3727$ line, as it is has two unsolved components separated in wavelength by 2.8 \AA. 

We further normalized the spectrum and determined EW of all the identified lines. Uncertainties were estimated from the EW and rms of the continuum according to the equation from \citet{1994ApJ...437..239G}:
\begin{equation}
\label{flux_error}
 \sigma _{i} = \sigma _{c} \sqrt{N+\frac{W _{i}}{\Delta }}
\end{equation}
where 
$\sigma _{i}$ is the error of the flux of the emission line, $\sigma _{c}$ is the {\it rms} measured from the local continuum, {\it N} is the measured line profile width in pixels, {\it $W _{i}$} is the absolute value of the line EW in  \AA, and {\it $\Delta$} is the spectral dispersion in \AA/pixel.
The observed flux measurements and the related parameters are listed in Table\,\ref{tab:emission_line_flux}, without any reddening corrections. The identified emission lines are those commonly seen in emission line galaxies, such as Green Peas (GPs) \citep{2012ApJ...749..185A}. 

The \Siii lines listed at the bottom of Table\,\ref{tab:emission_line_flux} were measured from the NIR spectrum which was taken on 2012 July 9 with the Telescopio Nazionale Galileo (TNG) + Near Infrared Camera Spectrometer (NICS) and previously presented in \cite{2013Natur.502..346N} (see Fig.\,\ref{fig:12dam_host_spec}). We fixed the width of FWHM of \Siii $\lambda9069$ line as the same velocity of \Siii $\lambda9532$. The flux ratio of $\lambda\lambda9532/9069$ is $\sim 2.85$. This is slightly different to the theoretical value of 2.44 expected in H\,{\sc ii} regions \citep{1982MNRAS.199.1025M}. However given the weakness of the features, the strong continuum and uncertainty in its placement, and the telluric absorption located around the \Siii wavelength range this ratio is reasonably consistent with the theoretical one. There was no re-scaling applied to match the NIR and the optical spectra due to no overlap in between.

Finally, we measured the host interstellar medium (ISM) absorption of the \Mgii $\lambda\lambda$2796/2803 doublet in the WHT spectrum of 2012 May 25 at the SN phase: $-13.6$d from the peak \citep{2013Natur.502..346N}. The rest-frame EWs are $2.11\pm0.27$ and $2.01\pm0.28$ \AA\, respectively, giving a ratio W$_{2796}$/W$_{2803}$ = 1.05. 
See section\,\ref{sec:metallicity} for further discussion of these values.

\begin{table*}
 \centering
 \begin{minipage}{140mm}
  \caption{Observed emission lines of the host of PTF12dam. The main optical spectrum was taken from the WHT on 2013 March 30. The lines highlighted in bold have been used to calculate the elemental abundances. We measured the gaussian line profiles within the {\sc procspec} task for strong lines (EW $\geq$ 10 \AA), and fixed the FWHM for weak lines (EW $<$ 10 \AA) in principle, except some lines marked with a $^{\ast}$ symbol (see details in the section \ref{sec:host_galaxy_spectroscopy}). 
We measured \Siii $\lambda9069$ and $\lambda9532$ lines from the NIR spectrum taken on 9 July 2012 with the TNG + NICS \citep{2013Natur.502..346N}. 
However, there is no overlap wavelength region between the main optical and NIR spectra. Hence there is no re-scaling process for the NIR spectrum.
The luminosity was derived from line flux, considering the luminosity distance of 481.1 Mpc, without any reddening correction.}
\label{tab:emission_line_flux}
  \begin{tabular}[t]{llllllll}
  \hline
Line & $\lambda$ & Observed flux & Error & RMS & EW & FWHM & Luminosity\\
 & (\AA) & (erg s$^{-1}$ cm$^{-2}$) & (erg s$^{-1}$ cm$^{-2}$) & (erg s$^{-1}$ cm$^{-2}$) & ($\AA$)  & ($\AA$) & (erg s$^{-1}$) \\
\hline
\Hei &3188 & $1.74\times10^{-16}$ & $4.16\times10^{-17}$ & $1.01\times10^{-17}$ & 1.82	& 3.17 & $4.84\times10^{39}$ \\
H16 + \Hei &3704 		& $7.89\times10^{-17}$ & $1.10\times10^{-17}$ & $3.64\times10^{-18}$ & 1.04		& 3.68 & $2.19\times10^{39}$ \\
{\bf \Oii} &{\bf 3727} & $\mathbf{9.50\times10^{-15}}$  & $\mathbf{4.74\times10^{-17}} $& $\mathbf{3.64\times10^{-18}} $& {\bf 132.49} & {\bf 4.08} & $\mathbf{2.64\times10^{41}} $\\	
H12 &3750 				& $1.55\times10^{-16}$ & $9.31\times10^{-18}$ & $2.63\times10^{-18}$ & 2.17	& 3.72 & $4.31\times10^{39}$ \\
H11 &3770 				& $2.09\times10^{-16}$ & $1.34\times10^{-17}$ & $3.99\times10^{-18}$ & 2.84	& 3.74 & $5.80\times10^{39}$ \\
H10 &3798 				& $2.49\times10^{-16}$ & $1.86\times10^{-17}$ & $5.16\times10^{-18}$ & 3.40	& 3.77 & $6.91\times10^{39}$ \\
H9 &3835 				& $3.38\times10^{-16}$ & $1.96\times10^{-17}$ & $4.83\times10^{-18}$ & 4.85	& 3.81 & $9.40\times10^{39}$ \\
{\bf \Neiii} & {\bf 3868} 		& $\mathbf{2.96\times10^{-15}}$ & $\mathbf{4.27\times10^{-17}}$ & $\mathbf{5.56\times10^{-18}}$ & {\bf 41.92}  & {\bf 4.02} & $\mathbf{8.22\times10^{40}}$ \\
\Hei + H8 &3889 		& $1.05\times10^{-15}$ & $2.96\times10^{-17}$ & $5.61\times10^{-18}$ & 14.75 & 3.88 & $2.91\times10^{40}$ \\
\Neiii + H7 &3968 		& $1.76\times10^{-15}$ & $2.40\times10^{-17}$ & $3.78\times10^{-18}$ & 23.46 & 5.47 & $4.87\times10^{40}$ \\
\Nii + \Hei &4026 		& $8.69\times10^{-17}$ & $1.25\times10^{-17}$ & $4.57\times10^{-18}$ & 1.14	& 4.00 & $2.41\times10^{39}$ \\
\Sii &4069 				& $1.30\times10^{-16}$ & $1.30\times10^{-17}$ & $4.10\times10^{-18}$ & 1.79	& 4.04 & $3.62\times10^{39}$ \\
{\bf \hd} & {\bf 4102} 		& $\mathbf{1.54\times10^{-15}}$ & $\mathbf{2.46\times10^{-17}}$ & $\mathbf{4.07\times10^{-18}}$ & {\bf 22.77}  & {\bf 3.91} & $\mathbf{4.28\times10^{40}}$ \\
{\bf \hg} & {\bf 4340} 		& $\mathbf{2.74\times10^{-15}}$ & $\mathbf{2.60\times10^{-17}}$ & $\mathbf{3.30\times10^{-18}}$ & {\bf 43.08}  & {\bf 3.93} & $\mathbf{7.61\times10^{40}}$ \\
{\bf \Oiii}  & {\bf 4363}	 	& $\mathbf{5.32\times10^{-16}}$ & $\mathbf{1.23\times10^{-17}}$ & $\mathbf{2.79\times10^{-18}}$ & {\bf 8.23}	 & {\bf 4.33} & $\mathbf{1.48\times10^{40}}$ \\
\Hei &4471 				& $2.16\times10^{-16}$ & $1.76\times10^{-17}$ & $4.71\times10^{-18}$ & 3.41	& 4.44 & $5.99\times10^{39}$ \\
\Feiii &4658 			& $1.12\times10^{-16}$ & $1.99\times10^{-17}$ & $5.51\times10^{-18}$ & 1.76	&$5.86^{\ast}$ & $3.10\times10^{39}$ \\
\Heii &4686 			& $3.85\times10^{-17}$ & $9.35\times10^{-18}$ & $3.61\times10^{-18}$ & 0.62	& 4.65 & $1.07\times10^{39}$ \\
\Feiii &4702 			& $3.61\times10^{-17}$ & $7.75\times10^{-18}$ & $2.79\times10^{-18}$ & 0.60	& 4.67 & $1.00\times10^{39}$ \\
\Ariv + \Hei &4713 		& $8.93\times10^{-17}$ & $1.16\times10^{-17}$ & $3.53\times10^{-18}$ & 1.50	& 4.68 & $2.48\times10^{39}$ \\
{\bf \hb} & {\bf 4861} 		& $\mathbf{6.19\times10^{-15}}$ & $\mathbf{3.54\times10^{-17}}$ & $\mathbf{2.84\times10^{-18}}$ & {\bf 117.91} & {\bf 4.97} & $\mathbf{1.72\times10^{41}}$ \\
\Hei &4921 				& $6.85\times10^{-17}$ & $1.18\times10^{-17}$ & $4.03\times10^{-18}$ & 1.34	& 4.89 & $1.90\times10^{39}$ \\
{\bf \Oiii}  & {\bf 4959} 		& $\mathbf{1.26\times10^{-14}}$ & $\mathbf{5.82\times10^{-17}}$ & $\mathbf{3.46\times10^{-18}}$ & {\bf 228.19} & {\bf 4.86} & $\mathbf{3.49\times10^{41}}$ \\
\Feiii &4986 			& $7.44\times10^{-17}$ & $1.40\times10^{-17}$ & $4.13\times10^{-18}$ & 1.36	& 4.95 & $2.07\times10^{39}$ \\
{\bf \Oiii} & {\bf 5007} 		& $\mathbf{3.66\times10^{-14}}$ & $\mathbf{1.12\times10^{-16}}$ & $\mathbf{4.13\times10^{-18}}$ & {\bf 623.15} & {\bf 4.88} & $\mathbf{1.02\times10^{42}}$ \\
\Ni &5199 				& $5.42\times10^{-17}$ & $7.16\times10^{-18}$ & $2.53\times10^{-18}$ & 0.98	& 5.16 & $1.51\times10^{39}$ \\
\Hei &5876 				& $7.54\times10^{-16}$ & $1.48\times10^{-17}$ & $2.29\times10^{-18}$ & 21.73 & 6.36 & $2.09\times10^{40}$ \\
\Oi &6300 				& $3.20\times10^{-16}$ & $7.35\times10^{-18}$ & $1.50\times10^{-18}$ & 7.88	& 6.26 & $8.88\times10^{39}$ \\
\Siii &6312 			& $9.25\times10^{-17}$ & $6.10\times10^{-18}$ & $1.50\times10^{-18}$ & 2.26	& 6.27 & $2.57\times10^{39}$ \\
\Oi  &6364 				& $9.04\times10^{-17}$ & $3.21\times10^{-18}$ & $1.10\times10^{-18}$ & 2.21	& 6.32 & $2.51\times10^{39}$ \\
{\bf \Nii} & {\bf 6548} 		& $\mathbf{2.34\times10^{-16}}$ & $\mathbf{7.85\times10^{-18}}$ & $\mathbf{1.91\times10^{-18}}$ & {\bf 6.96}	 & {\bf 6.50} & $\mathbf{6.49\times10^{39}}$ \\
{\bf \ha} & {\bf 6563}			& $\mathbf{1.92\times10^{-14}}$ &$\mathbf{5.03\times10^{-17}}$ & $\mathbf{1.91\times10^{-18}}$ & {\bf 583.63}  & {\bf 6.30} & $\mathbf{5.34\times10^{41}}$\\
{\bf \Nii} &{\bf 6583} 		& $\mathbf{6.99\times10^{-16}}$ & $\mathbf{1.27\times10^{-17}}$ & $\mathbf{1.91\times10^{-18}}$ & {\bf 21.85}  & $\mathbf{6.54^{\ast}}$ & $\mathbf{1.94\times10^{40}}$\\
\Hei &6678 				& $2.54\times10^{-16}$ & $1.39\times10^{-17}$ & $2.72\times10^{-18}$ & 8.84	& 6.63 & $7.05\times10^{39}$ \\
{\bf \Sii} & {\bf 6717} 		& $\mathbf{1.16\times10^{-15}}$ & $\mathbf{1.56\times10^{-17}}$ & $\mathbf{1.92\times10^{-18}}$ & {\bf 42.83}  & {\bf 6.76} & $\mathbf{3.22\times10^{40}}$\\
{\bf \Sii} & {\bf 6731} 		& $\mathbf{8.82\times10^{-16}}$ & $\mathbf{1.42\times10^{-17}}$ & $\mathbf{1.92\times10^{-18}}$ & {\bf 33.14}  & {\bf 6.86} & $\mathbf{2.45\times10^{40}}$ \\
\Hei &7065 				& $2.87\times10^{-16}$ & $1.38\times10^{-17}$ & $2.68\times10^{-18}$ & 9.83	& 7.01 & $7.97\times10^{39}$ \\
\Ariii &7136 			& $5.43\times10^{-16}$ & $1.93\times10^{-17}$ & $3.03\times10^{-18}$ & 18.59 & 6.37 & $1.51\times10^{40}$ \\
{\bf \Oii} & {\bf 7320} 			& $\mathbf{2.78\times10^{-16}}$ & $\mathbf{7.29\times10^{-18}}$ & $\mathbf{1.57\times10^{-18}}$  & {\bf 7.51}	 & {\bf 7.27} & $\mathbf{7.73\times10^{39}}$ \\
{\bf \Oii} & {\bf 7330} 			& $\mathbf{1.97\times10^{-16}}$ &$\mathbf{6.87\times10^{-18}}$ &  $\mathbf{1.57\times10^{-18}}$ & {\bf 5.38}	 & {\bf 7.28} & $\mathbf{5.47\times10^{39}}$ \\
\hline
{\bf \Siii}&{\bf 9069} 			& $\mathbf{7.00\times10^{-16}}$ & $\mathbf{2.33\times10^{-17}}$ &  $\mathbf{7.41\times10^{-18}}$ &  {\bf 5.38} &  {\bf 17.55} & $\mathbf{1.94\times10^{40}}$ \\
{\bf \Siii}&{\bf 9532} 			& $\mathbf{2.00\times10^{-15}}$ & $\mathbf{2.21\times10^{-17}}$ & $\mathbf{5.33\times10^{-18}}$ &  {\bf 17.28} & {\bf 18.44} &  $\mathbf{5.55\times10^{40}}$ \\
\hline
\end{tabular}
\end{minipage}
\end{table*}

\subsection{Late time monitoring of SLSN PTF12dam}
\label{sec:late_time_monitoring_of_12dam}

The rise time of the early phase lightcurves presented in \cite{2013Natur.502..346N} suggested that pair-instability models could not consistently fit the data. 
The lightcurves were better fit with simple magnetar engine models. However, continuous photometric follow-up to as late times as possible is essential to test this conclusion. 
To monitor the flux of PTF12dam beyond the timescales presented in \cite{2013Natur.502..346N}, we imaged the source in optical and NIR filters on multiple epochs from +230d to +550d, and additionally gathered a SN spectrum at +509d after peak (rest-frame) with the 10.4-m Gran
Telescopio CANARIAS (GTC). Although this spectrum was taken much later than the WHT + ISIS 
spectrum it was not suitable for a metallicity analysis as it has much lower resolution, narrower wavelength coverage, and the
absolute flux calibration was not based on such careful observing procedures as employed at the 
WHT. All the spectroscopic data used in this paper are summarised in 
Table\,\ref{tab:spectroscopy}. The log of photometric observations is 
in Table\,\ref{tab:phot}. 
We defined the epoch of zero days as the date of {\it r}-band maximum on 2012 June 10 (MJD = 56088), rather than referring to the estimated date of SN explosion (MJD = 56017) from the magnetar model fit \citep{2013Natur.502..346N}. 
All phases quoted here have been corrected for cosmological time dilation using the observed redshift $z = 0.107$ and are thus in the rest-frame with respect to this peak magnitude date. 

\begin{table*}
 \centering
 \begin{minipage}{190mm}
  \caption{Late-time photometry of PTF12dam and its host galaxy. The phase is in the SN rest-frame and it is days after {\it r}-band maximum (MJD = 56088). The ``$>$'' denotes that a detection was not made and the magnitude provided is the 3$\sigma$ detection limit. The ``template" images are assumed to be the host galaxy only with no SN flux contribution. Optical {\it gri} magnitudes are in the AB system and the NIR {\it JHK$_{s}$} magnitudes are in the 2MASS Vega system.}
\label{tab:phot}
\begin{tabular}[t]{lllllllllll}
\hline
 Date & MJD & Phase & {\it g} &  {\it r} &  {\it i} & {\it J} & {\it H} & {\it $K_{s}$} &Telescope+Instrument\\
\hline
2013 Feb 20 & 56343.25 & 230.58& & & & $19.18\pm0.19$ & $18.81\pm0.14$ & $17.90\pm0.19$ & NOT + NOTCam\\
2013 Mar 11 & 56362.26 & 247.75 & $20.86\pm0.09$ & $20.36\pm0.04$ & $20.83\pm0.11$ &&&& LT + RATCam\\
2013 Mar 23 &56374.06	&258.41& & & & $19.73\pm0.16$ & $19.16\pm0.13$ & $19.03\pm0.18$ & NOT + NOTCam \\
2013 Apr 23 & 56405.06&286.41 & $>20.9$ & $>21.4$ & $>19.8$ &&&& LT + RATCam\\
2013 Apr 26 & 56408.11	&289.17 & & & & $19.44\pm0.12$ & $19.61\pm0.21$ & $>19.3$ & NOT + NOTCam \\
2013 May 12& 56424.05	&303.57 & $21.36\pm0.09$ & & &&&& LT + RATCam\\
2013 May 20& 56432.93	&311.59 & & $21.73\pm0.09$ & &&&& LT + RATCam\\
2013 Jun 09 &56452.96&329.68 & $22.25\pm0.07$ & $22.32\pm0.04$ & $22.12\pm0.05$ &&&& WHT + ACAM \\
2013 Jun 24 & 56467.03	&342.39 & & &  & $20.12\pm0.16$& $>18.7$ &  & NOT + NOTCam \\ 
2013 Aug 25& 56529.90	&399.19& $23.97\pm0.39$ & $>24.1$ & $23.33\pm0.16$ &&&& WHT + ACAM\\
\hline
Host Template & & & & & & & & & \\
\hline
2014 Jan 22 & 56679.11 & 533.97 & $19.50\pm0.08$ & $19.11\pm0.04$ & $18.81\pm0.03$ & & & & WHT + ACAM\\
2014 Feb 09 & 56697.48 & 550.57 & $19.33\pm0.04$ & $19.00\pm0.03$ & $18.85\pm0.03$ & & & & LT+ RATCam\\
2014 Feb 13 & 56701.17 & 553.90 &  & & &$18.37\pm0.15$ & $17.82\pm0.13$& $17.16\pm0.09$ & NOT + NOTCam\\
\hline
2003 Feb 11 & 52681.47 & pre-explosion & $19.36\pm0.02$ & $19.20\pm0.02$ & $18.83\pm0.03$ &&&& SDSS DR9 \\
\hline 
\end{tabular}
\end{minipage}
\end{table*}

\subsubsection[]{Late-time photometric evolution of PTF12dam}
\label{sec:late_time_photometric_evolution_of_12dam}

To cover the NIR ({\it JHK$_{s}$} bands), four epochs were taken by the NOT + NOTCam 
between +230d and +342d (for details of NOTCam, see section\,\ref{sec:host_galaxy_photometry}). 
We obtained deep images at phase +554d which we assumed contained host galaxy flux only
and negligible SN flux (as discussed in section\,\ref{sec:host_galaxy_photometry}). 
These images were used as templates and subtracted from the 
four previous epochs. To do this, we aligned the images using the {\sc geomap} and {\sc geotran} tasks within {\sc iraf} package and employed the image subtraction package, High Order Transform of PSF ANd Template Subtraction ({\sc HOTPANTS})\footnote{http://www.astro.washington.edu/users/becker/v2.0/hotpants.html}. The FWHM of the point-spread-function (PSF) of the SN flux in these 
difference images was compared with the FWHM of surrounding stars. We found 
it to have similar values indicating that we were recovering a point-like source. 

Magnitude measurements of the SN were carried out using aperture photometry with the {\sc iraf/daophot} task. We let the aperture size vary until we were confident that it encompassed the whole SN flux but avoided any negative pixels appearing in some subtraction images.
The zeropoint was set as discussed in section\,\ref{sec:host_galaxy_photometry}. 
The SN magnitudes in these 4 epochs are listed in Table\,\ref{tab:phot}. 
In cases where the SN was undetected, we determined a $3\sigma$ limiting magnitude. 
We added 10 fake stars, of varying magnitude, within 1 arcmin radius around the SN position. We measured their magnitudes and uncertainties and then determined the $3\sigma$ detection limit to be when the standard deviation of the photometry of these fake stars was 0.3 mag. These detection limits are listed in Table\,\ref{tab:phot}. For consistency, we double checked that the detected SN magnitudes were brighter than the estimated detection limit of each frame.

\begin{figure}
\includegraphics[angle=0,width=\linewidth]{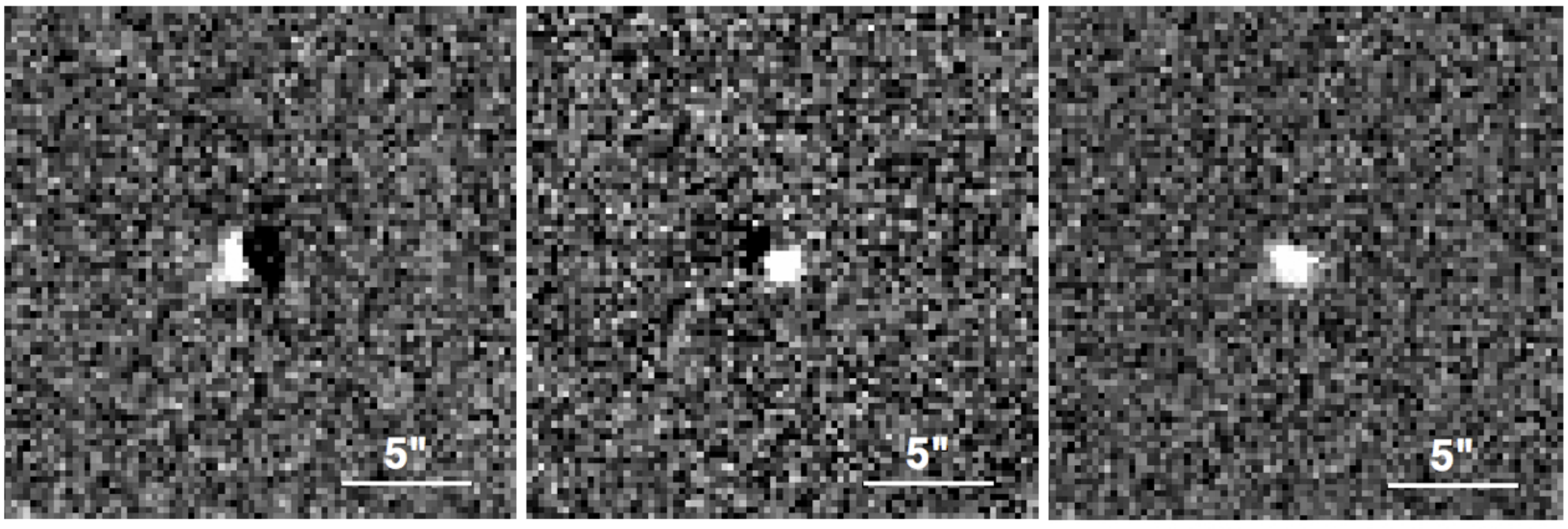}
\caption{Image subtraction examples with the {\it i}-band images of latest epochs of PTF12dam. All images were taken from WHT + ACAM. {\it Left:} 20130825 $-$ 20140122. {\it Middle:} 20130609 $-$ 130825. {\it Right: 20130609 $-$ 20140122}.}
\label{fig:12dam_sub}
\end{figure}

The optical frames ({\it gri} bands) were taken with two different instrumental configurations: 4 epochs between +247d and +312d from the LT + RATCam, and 2 later epochs (after +329d) from the WHT + ACAM. We also took deep host images after +530d using the same instruments to use
as host galaxy templates and the observation log for these data is listed in Table\,\ref{tab:phot}. 
We assumed the LT images at +534d and the WHT images at +550d were dominated by 
the host galaxy flux and could be used as pure host templates. 
Although we can detect some broad emission line flux from PTF12dam in the 
+509d spectrum (see section\,\ref{sec:GTC_spec}), the overall SN flux in the 
broad band filters was negligible. For example, in the {\it i} band, the SN flux is less than $\sim$ 1\% of the host flux ($i_{\rm SN} \sim$ 25 compared with $i_{\rm host} \sim$ 19).

The RATCam camera consists of a $2048 \times 2048$ EEV CCD giving a field of view of 4.6 arcmin and 0\farcs135 per pixel sampling. This camera was formally decommissioned in February 2014 and we gathered images of PTF12dam (4 epochs) and host template images through the filters {\it gri} before this replacement. 
The images were processed through the standard LT data reduction pipeline (bias-subtracted and flat-fielded). As with the NIR data, we aligned the images first using {\sc iraf/geomap} and {\sc iraf/geotran} tasks, and then subtracted SN images with host templates through {\sc HOTPANTS}. Photometric flux measurements were performed using the {\sc iraf/daophot} task and aperture photometry. 
We varied the aperture size until all of the SN flux was covered, then applied the same size for measuring the SN and standards.
The zeropoint was derived from 5 standard stars within the SDSS DR9 catalog. Table\,\ref{tab:phot} lists the photometry results. Due to poor weather conditions the LT images at +286d yielded 
limiting magnitudes only. 

The ACAM instrument on the WHT (camera specifications described in section\,\ref{sec:host_galaxy_photometry})
provided deeper images at later times to allow us to recover the faint SN flux
signal within the bright host galaxy. This was challenging as the SN flux was around
1\% of the host galaxy flux and the position of the PTF12dam is close to the brightest 
pixels in compact host. The analysis process as described for the RATCam data
was used for the ACAM images. In this case there were 10 SDSS DR9 reference stars around the SN to calibrate the ACAM images. We again employed {\sc HOTPANTS} to carry out 
target frame minus template subtractions, but the subtraction process was complicated by non-circular
PSFs on some of the later frames. The last epoch for which we attempt to measure SN flux in 
(+400d taken on 2013 August 25) proved to be challenging. 
The {\it i}-band image taken on this date (which we will call the 20130825 image for brevity) 
had a relatively large ellipticity, with a FWHM major/minor axes ratio of 1.04/0.86\arcsec. 
The {\it i}-band host galaxy template image, taken on 2014 January 22 
(which we will call 20140122) has a more circular PSF, with axes ratios of 1.00/0.95\arcsec. 
We ran {\sc HOTPANTS} with two different convolution directions, but in each of these
cases it resulted in an attempted deconvolution along one of the minor or major axes directions. 
This produced poor quality difference images with visible dipole residuals (see left panel of Fig.\,\ref{fig:12dam_sub}). Despite various attempts, no satisfactory result could be achieved to carry our the 20130825 $-$ 20140122 image subtraction. 

We thus employed an alternative approach. We subtracted the 20130825 and 
20140122 images from the 20130609 image, which has a smaller and circular FWHM (major/minor axes measurements of 0.77/0.72\arcsec). This produced significantly better
difference images (see middle and right panel of Fig.\,\ref{fig:12dam_sub}), with no obvious residuals. 
The difference between the flux measured in the image 20130609 $-$ 20140122 and that measured in 20130609 $-$ 20130825 is then representative for our required epoch of 20130825 $-$ 20140122. 
The {\it i}-band magnitude of the SN source in the difference image 20130609 $-$ 20140121 is $22.12\pm0.05$, and in 20130609 $-$ 20130825 difference image it is $22.56\pm0.06$, hence the magnitude of the SN in the 20130825 $-$ 20140122 is $i=23.33\pm0.16$ mag. 
The same analysis method and calculation was applied to the {\it g}-band data to give a final magnitude of $23.97\pm0.39$. There was no visible source detected in the {\it r}-band subtractions, and we set an upper limit of $r > 24.1$ mag.

\begin{figure}
\includegraphics[angle=0,width=\linewidth]{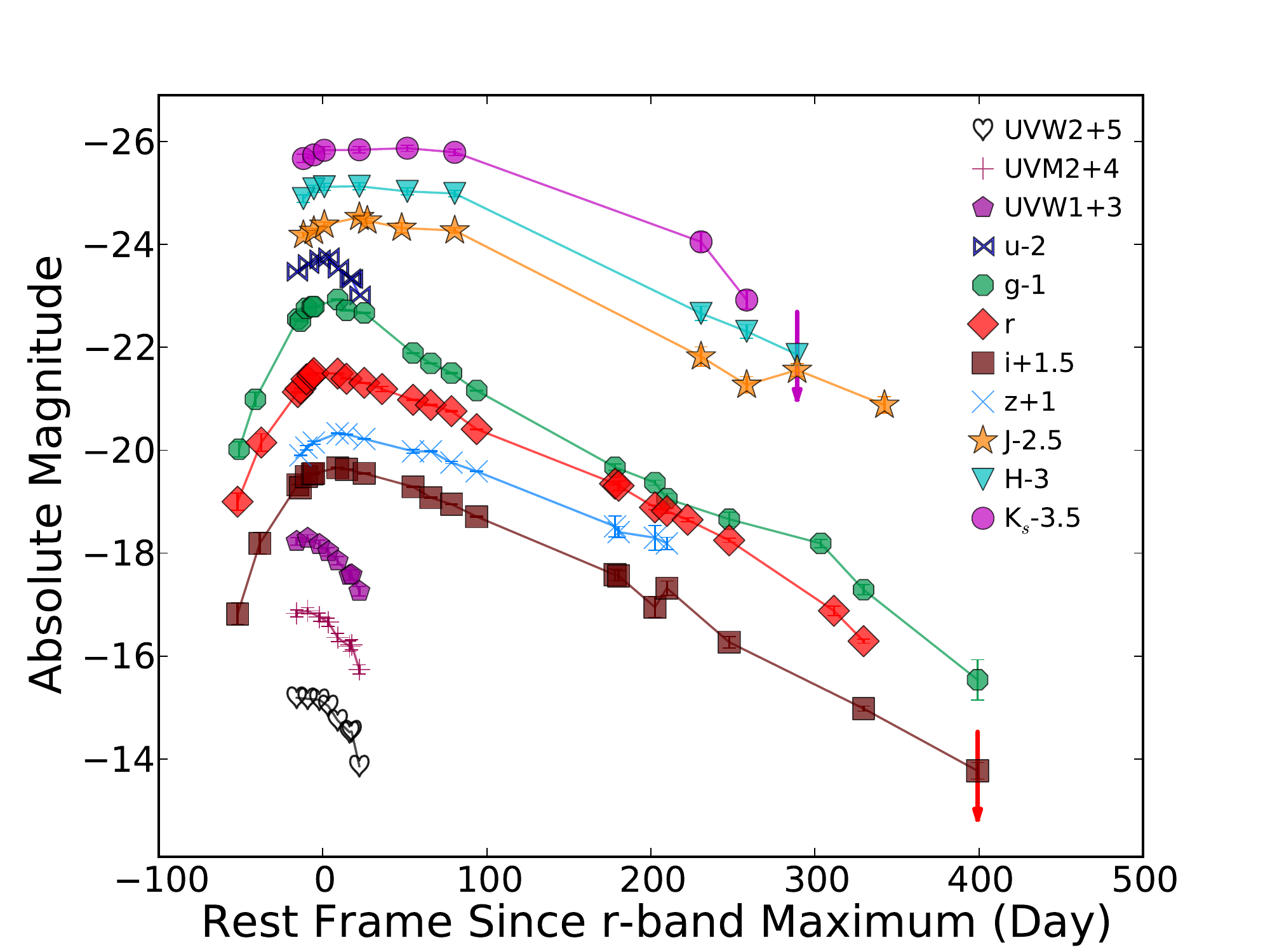}
\caption{Absolute lightcurves of PTF12dam. Photometric points after +230d were based on this work applying for the proper host subtraction; before +230d were taken from \citet{2013Natur.502..346N}. Upper limits are shown with arrows of {\it r} and {\it K$_{s}$} bands at the last epochs. All points have been corrected for extinction, and {\it griz} magnitudes have also been {\it K}-corrected.}
\label{fig:12dam_abs_lc}
\end{figure}

The results are summarised in Table\,\ref{tab:phot} and the absolute lightcurves, from UV to NIR, are plotted in Fig.\,\ref{fig:12dam_abs_lc}. The early phases (before +230d) were taken from \citet{2013Natur.502..346N}, and the late-time magnitudes are the results of the template subtractions carried out here. 
Due to the lack of late-time spectra of PTF12dam, we adopted the {\it K}-correction values from the final SN spectra around +200d. There were {\it K$_{g}$}=0.24, {\it K$_{r}$}=0.04, and zero for {\it i} and {\it JHK$_{s}$}. 
The foreground ({\it A$_{v}$}=0.037) and internal dust extinction ({\it A$_{v}$}=0.2) are also corrected for. We discuss the estimate of host internal dust effect in the section \ref{sec:galaxy_size}.

\subsubsection{Late-time spectroscopy of PTF12dam}
\label{sec:GTC_spec}
We took a spectrum of PTF12dam and its host galaxy with the 10.4-m Gran Telescopio CANARIAS (GTC) + Optical System for Imaging and low-Intermediate-Resolution Integrated Spectroscopy (OSIRIS) on 2013 December 25. The R500R grating was used (dispersion = 4.88 \AA/pixel) with a central wavelength 7165 \AA\ to cover the range from 4900 to 9200 \AA. We used a 1\farcs0 slit to obtain three spectra of 2400 sec exposures each. All frames were reduced (overscan corrected, bias-subtracted and flat-fielded) though the {\sc IRAF} package. The spectra were extracted using the standard {\sc IRAF} routines, wavelength-calibrated by identifying lines of HgArXeNe arcs, and flux-calibrated by comparing with a spectroscopic standard star GD248.
Telluric features were identified in reduced standard star spectra and subsequently removed from the SN spectra. The spectroscopic data are summarised in Table\,\ref{tab:spectroscopy}. 

 At this epoch, the SN was +508.6d after the peak (rest-frame) which means that the spectrum 
was dominated by emission lines and continuum from the host galaxy. Hence a method to 
retrieve any residual, broad lines in the SN nebular phase was required 
\citep[e.g. see the nebular spectrum of SN~2007bi from][]{2009Natur.462..624G}. 
We first scaled the flux of the GTC spectrum to match the host galaxy photometry with SDSS magnitudes. Then we applied a correction for the Milky Way reddening, shifted the spectrum to the rest frame, and applied an internal dust extinction correction (for all values used here, see Table.\,\ref{tab:property}).

\begin{figure}
\includegraphics[angle=0,width=\linewidth]{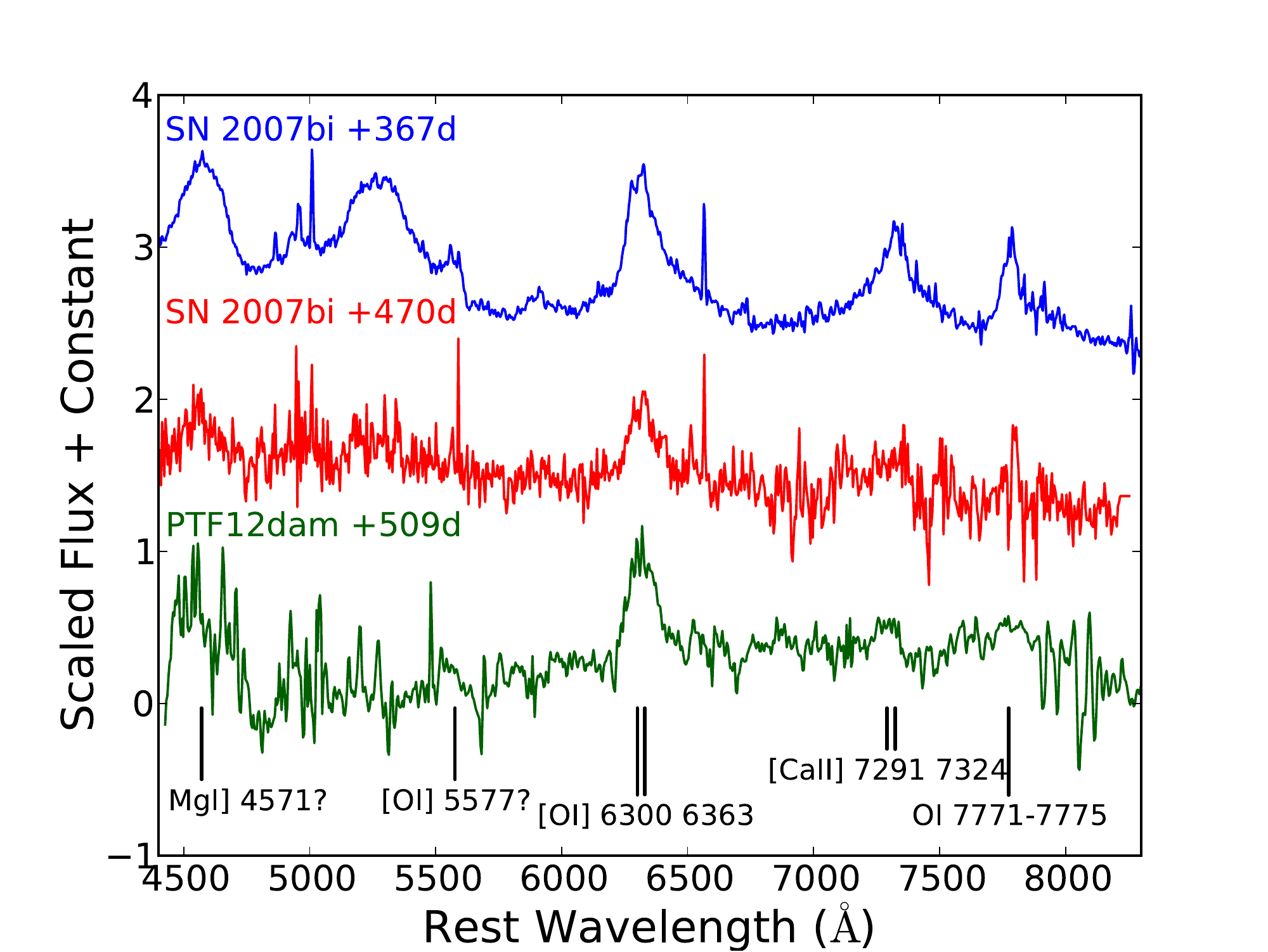}
\caption{Spectrum of the host galaxy of PTF12dam at +508.6d taken by GTC. 
Comparison with the late-time spectra of SN~2007bi from the WISeREP database \citep{2012PASP..124..668Y}, which were published on \citet{2010A&A...512A..70Y, 2012Sci...337..927G}. Both wavelength and phases are stated in the SN rest-frame.}
\label{fig:12dam_GTC_spec}
\end{figure}

We subtracted a galaxy continuum model, using the same method as in \cite{2013Natur.502..346N}. 
The population synthesis code
$starburst99$ \citep{1999ApJS..123....3L} was used to produce a model of a 30 Myr stellar population with continuous star formation, a metallicity of 0.05 \zsol and a Salpeter IMF \citep[further details in][]{2013Natur.502..346N}. 
In order to be consistent with the previous and published result, we adopted the same metallicity model of the galaxy continuum, although the gaseous metallicity we measured is higher ($\sim0.2$ \zsun).
We scaled the galaxy model to match the observed continuum in 
the observed spectrum, since we know the observed spectrum must be dominated by the host and the 
overall SN flux is around 1-2\%. This model was subtracted from the observed
spectrum. Finally we removed the host galaxy emission lines by fitting them with gaussian profiles and subtraction. This process left noise residuals at the positions of the strong lines, 
which we smoothed over with interpolation. 
Since we have a measurement of the SN flux in the {\it i} band from image subtraction, we scaled the
resultant spectrum with a constant multiplicative factor so that it had a synthetic {\it i}-band
magnitude of {\it i} = 25.0 (in the observer frame).
 The final spectrum, which is our best attempt at recovering the uncontaminated 
SN flux of PTF12dam (at an epoch of +509d) is shown in Fig.\,\ref{fig:12dam_GTC_spec}. 
While this is understandably noisy, and the absolute flux of the SN continuum is uncertain, we do 
clearly detect the broad \Oi $\lambda\lambda$6300,6363 emission line feature from PTF12dam. 
The relative strength of this line is quite similar to that detected in SN~2007bi at +470d. 
The host galaxy of SN~2007bi is some 3.4$^{m}$ fainter than that of PTF12dam, hence
the nebular spectrum of \cite{2009Natur.462..624G} suffers much less host contamination. 
The identification of other features is more uncertain, but we label these possible detections 
in Fig.\,\ref{fig:12dam_GTC_spec}.

To quantitatively compare these emission lines to those of SN~2007bi, we fitted gaussian profiles
to the broad SN features to estimate flux, FWHM and EWs (again using our custom built
{\sc IDL} suite of programmes {\sc procspec}). The largest uncertainty in measuring the line flux comes
from the positioning and definition of the continuum. We calculated values of the line flux
and EWs from various reasonable continuum choices and different polynomial fitting functions (from order 1 to 3). 
The average values and the standard deviation of these are quoted in Table\,\ref{tab:sn_line_flux}. 
The continuum positioning and choices were somewhat subjective and given the model subtraction and scaling, these
flux and uncertainties should be treated with some caution. They are meant to illustrate that that 
we do detect the broad oxygen nebular lines in PTF12dam and give an approximate line flux. 

The expansion velocities in Table\,\ref{tab:sn_line_flux} are the velocity values for the FWHM of the features. 
For the blended lines, we fitted single gaussians and assumed that the two components were roughly equal in 
strength. The velocity estimated is a simple estimate of a deconvolved component of the doublet, assuming
the intrinsic width of the component is just the separation between the lines. 
Since the GTC spectrum of PTF12dam is quite similar in morphology with the late-time spectra of SN~2007bi, we also measured the above properties for the \Oi $\lambda\lambda$ 6300,6363 line feature in SN~2007bi.
All nebular line measurements are listed in Table\,\ref{tab:sn_line_flux}. The line fluxes for SN~2007bi are significantly higher than those for PTF12dam, and the EW a factor of nearly two lower. This suggests that there is 
some host galaxy contamination in the last SN~2007bi spectrum of 
\cite{2009Natur.462..624G}. With the new measurements we now have of the host of SN~2007bi (see section\,\ref{sec:host_galaxy_photometry}) this could 
be quantitatively addressed in future work.

\begin{table*}
 \centering
 \begin{minipage}{140mm}
  \caption{Flux measurements of the broad SN lines of PTF12dam in the GTC spectrum taken at +509d. 
The \Oi line flux of SN~2007bi are reported for comparison.}
\label{tab:sn_line_flux}
  \begin{tabular}[t]{llllllll}
  \hline
SN Name & Line & $\lambda$ & Flux $\pm$ Error & EW & FWHM & Velocity & Luminosity $\pm$ Error \\
 & & (\AA) & (erg s$^{-1}$ cm$^{-2}$)  & ($\AA$)  & ($\AA$) & (km s$^{-1}$) & (erg s$^{-1}$) \\
\hline
PTF12dam (+509d) & \Oi &5577 &  $7.0\pm0.5\times10^{-18}$ & 187 & 74 & $\sim$ 4000 &  $1.9\pm0.2\times10^{38}$  \\
& {\bf \Oi} &{\bf 6300 6363}&  $\mathbf{4.6\pm0.3\times10^{-17}}$ & {\bf 332} & {\bf 137} & {\bf $\sim$ 5800} &  $\mathbf{1.3\pm0.1\times10^{39}}$ \\
& \Caii & 7291 7324 & $1.1\pm0.1\times10^{-17}$ & 71 & 102 & $\sim$ 4000 &  $2.9\pm0.3\times10^{38}$ \\
& \Oinir & 7771-7775 & $1.2\pm0.1\times10^{-17}$ & 78 & 109 & $\sim$ 4200 &  $3.3\pm0.4\times10^{38}$ \\ 
SN~2007bi (+470d) & {\bf \Oi} &{\bf 6300 6363} &  $\mathbf{2.4\pm0.3\times10^{-16}}$  & {\bf 190} & {\bf 143} & {\bf $\sim$ 6100} &  $\mathbf{9.5\pm1.0\times10^{39}}$\\
SN~2007bi (+367d) & {\bf \Oi} &{\bf 6300 6363} &  $\mathbf{6.0\pm0.4\times10^{-16}}$ & {\bf 358} & {\bf 182} &  {\bf $\sim$ 8100} & $\mathbf{2.4\pm0.2\times10^{40}}$\\
\hline
\end{tabular}
\end{minipage}
\end{table*}

\section[]{Host galaxy properties}
\label{sec:host_galaxy_properties}

\subsection{Galaxy size, extinction corrections and luminosity}
\label{sec:galaxy_size}

The profile of the host galaxy is noticeably broader than the stellar PSF, hence we can estimate a physical diameter of the extended source, assuming the relation (galaxy observed FWHM)$^{2}$ = (PSF FWHM)$^{2}$ + (intrinsic galaxy FWHM)$^{2}$. 
We measured the FWHM of both the host galaxy (1.48\arcsec) and the average (1.09\arcsec) of 10 reference stars within a 5 arcmin radius around the host in the SDSS {\it r}-band images taken on 2003 February 11. This provides a physical diameter of 1.9 kpc at the angular size distance 392.6 Mpc for $z = 0.107$ \citep{2006PASP..118.1711W}, 
assuming a cosmology of H$_{0}=72$\,km\,s$^{-1}$\,Mpc$^{-1}$, $\Omega_{\rm M}=0.3$, $\Omega_{\rm \lambda}=0.7$ . 
We also applied the same process for the {\it r}-band image taken from the WHT on 2014 January 22. The FWHM of the PTF12dam host was found to be 1.52\arcsec, and the seeing was 1.22\arcsec. We calculated that the physical diameter of host to be 1.7 kpc, which is consistent with the result (1.9 kpc) from the SDSS pre-explosion image. Therefore, we adopted 1.9 kpc for the physical diameter of PTF12dam host as it was under a better seeing condition.

The absolute magnitude of the host is {\it M$_{g}$} = $-19.33\pm0.10$, 
after correcting for foreground extinction, {\it K}-correction, and internal dust extinction corrections. 
The methods for these corrections are described as follows. 
Firstly, in the optical, we adopted the apparent Petrosian magnitudes from the SDSS DR9 catalog \citep{2012ApJS..203...21A} of the host galaxy (pre-explosion); for the UV bands, we adopted magnitudes from the GALEX GR6 catalog (pre-explosion); for the NIR, we used our measurement of the host taken after +554d from the SN peak. All photometric values are reported in section\,\ref{sec:host_galaxy_photometry}.
And then we applied a Milky Way extinction correction \citep {2011ApJ...737..103S} for each bandpass based on the value of {\it A$_{V}$} = 0.03. 

Next we determined {\it K}-corrections of the host using two methods. The first was using the recipes of \citet{2010MNRAS.405.1409C} and \cite{2012MNRAS.419.1727C} with the SDSS colours. The second
was using our own {\it K}-correction code (Inserra et al. in prep.) with the input spectrum of the host. All the corrections were minimal (less than a few hundredths of a magnitude) apart from the {\it i} band, which we calculated from our own code to have a correction of $-0.06$ mag. We also applied a {\it K}-correction to the GALEX magnitudes of {\it FUV} = $-0.05$ mag \citep[from][]{2010MNRAS.405.1409C, 2012MNRAS.419.1727C}.

Moreover, we applied an internal dust extinction correction. 
We assumed a host extinction curve with {\it R$_{V}$} = 4.05 that may be more appropriate for star-forming galaxies \citep{2000ApJ...533..682C}. The median SFR of their star-forming galaxy sample is 7.2 \msun\,year$^{-1}$, which is comparable with the SFR of the host of PTF12dam (5.0 \msun\,year$^{-1}$). We then used the intrinsic line ratio of \ha/\hb = 2.86, \hg/\hb = 0.47 and \hd/\hb = 0.26 assuming case B recombination for $T\rm_{e}$ = 10000\,K and $n \rm_{e}$ = 100\,cm$^{-3}$ \citep{1989agna.book.....O, 2006agna.book.....O}, which implied an internal dust extinction of approximately {\it A$_{V}$} = 0.20. Finally the absolute magnitudes were then calculated after these three corrections, using the luminosity distance of 481.1 Mpc and are listed in Table\,\ref{tab:property}.

For comparison we also estimate here the equivalent values for the host of SN~2007bi. The \ha to \hb ratio
as published in the spectra of \cite{2010A&A...512A..70Y} implied a negligible internal dust extinction. Hence we only applied 
de-reddening for the foreground extinction from the \citet{2011ApJ...737..103S} Galactic dust map ({\it A$_{g} = 0.09$}) and 
then a {\it K}-correction ({\it g} = $-0.01$) from \citet {2010MNRAS.405.1409C, 2012MNRAS.419.1727C}. This resulted in absolute magnitudes (for a luminosity distance of 578.5 Mpc) of {\it M$_{g}$} = $-16.05\pm0.10$, 
{\it M$_{r}$} = $-16.49\pm0.06$, {\it M$_{i}$} = $-16.45\pm0.08$ and {\it M$_{z}$} = $-16.81\pm0.14$. The host of PTF12dam is some 3 magnitudes brighter than that of SN~2007bi. As shown above, this complicates the recovery of SN flux at late times. 

In addition, we measured the physical size of the host of SN~2007bi using the same method as employed above for PTF12dam. In the WHT {\it r}-band image on 2012 May 25, the seeing was 0.79\arcsec from the average of 7 stars within a 1 arcmin radius around the host, while the FWHM of the host was 1.25\arcsec. Based on the angular size distance of the host of 455.5 Mpc, the physical diameter (derived from the observed galaxy FWHM) of the host of SN~2007bi is estimated to be 2.1 kpc. 

\subsection{Galaxy stellar mass}
\label{sec:stellar_mass}

We employed the {\sc MAGPHYS} stellar population model program of \citet{2008MNRAS.388.1595D} to estimate the stellar mass from the observed photometric points across UV, optical and NIR bands. The program employs a library of stellar evolution and population synthesis models from \citet{2003MNRAS.344.1000B} and fits the luminosity of the host stellar population.

The PTF12dam host spectrum is dominated by strong nebular emission lines which are not included in the model galaxy spectra of {\sc MAGPHYS}. In order to estimate the appropriate broad band magnitudes to input into {\sc MAGPHYS}, the contribution of these emission lines need to be accounted for. 
We employed the WHT spectrum taken on 2013 March 30 to do this. One could simply remove the
narrow emission lines from this spectrum 
and calculate synthetic photometry of the resultant spectrum to 
estimate the contribution from the lines. However this spectrum also contains some flux of PTF12dam (at +266d rest-frame) as discussed in section\,\ref{sec:observational_data}. Therefore we carried out several experiments to estimate the 
emission line contamination to subtract from the SDSS pre-discovery host galaxy magnitudes. 
In the first instance, we attempted to remove the SN contribution by scaling the PTF12dam spectrum at +221d
by appropriate values estimated from the observed lightcurves (see Table\,\ref{tab:phot}) and subtracting this from the observed WHT spectrum.
Secondly, we removed the emission lines in the WHT +266d spectrum 
by fitting low order polynomials across the base of the line widths and replacing the lines with these
fits. We carried out synthetic photometry before and after removing the lines in the {\it gri} bands.
To a level of accuracy of 0.1 mag, we 
found corrections to the {\it gri} bands of 0.2, 0.2 and 0.7 mag respectively.
This large correction for the {\it i} band is due to the major contribution of redshifted \ha, whereas the 
strong [O{\sc iii}] lines fall in the {\it g}/{\it r}-band filter cross-over point.

\begin{figure}
\includegraphics[angle=0,width=\linewidth]{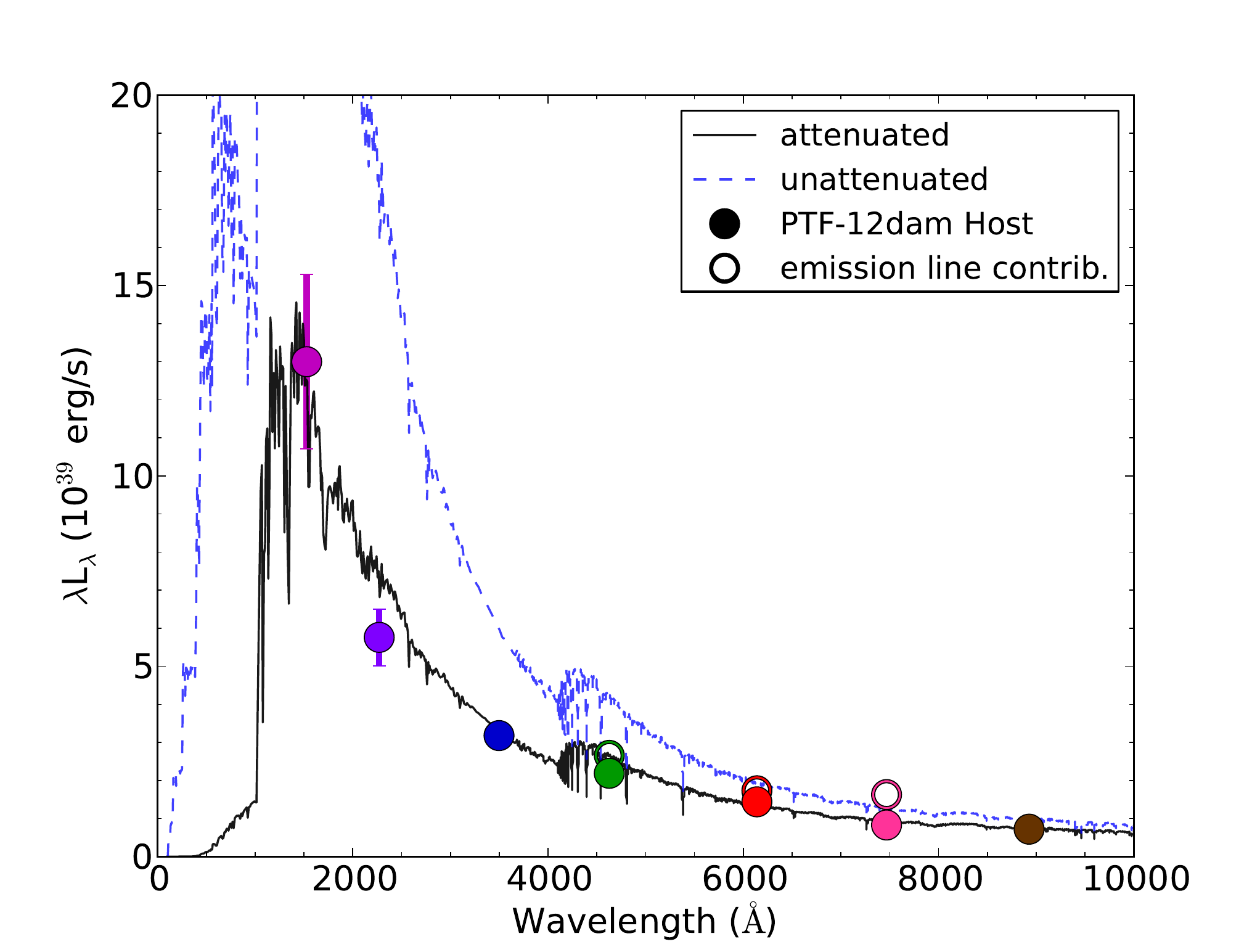}
\caption{The best-fit ($\chi^{2}=1.3$) SED of the host galaxy of PTF12dam from {\sc MAGPHYS} \citep{2008MNRAS.388.1595D}, with (black line) and without (blue line) an attenuation by internal dust. 
The solid circles show the photometry from the galaxy continuum component, in the order of GALEX {\it FUV}, {\it NUV} and SDSS {\it ugriz} filter regions. Since there are no strong emission lines in {\it FUV}, {\it NUV}, {\it u} and {\it z} bands, we only removed the emission line contributions in {\it gri} bands. The hollow circles show their original photometric values of the host including emission line flux.}
\label{fig:12dam_sed}
\end{figure}

We used these {\it gri} continuum magnitudes along with the SDSS observed {\it z}-band magnitude, the GALEX {\it FUV} and {\it NUV} magnitudes and our measured {\it JHK$_{s}$} magnitudes from the last epoch described in section\,\ref{sec:host_galaxy_photometry} as measurements of the galaxy continuum from the FUV to the NIR. For the bands other than {\it gri} we did not apply any emission line correction since the
galaxy spectra are dominated by continuum in the FUV and NIR and emission line contributions are not as strong as in the observed {\it gri} wavelength region. We used all of these bands as input to {\sc MAGPHYS} \citep{2008MNRAS.388.1595D}, which 
determines a spectral energy distribution (SED) of the best-fit model. This is simply the model with the lowest $\chi^2$, estimated after allowing the physical input parameters to vary. The SED of the best-fit model with $\chi^2=1.3$ is plotted in Fig.\,\ref{fig:12dam_sed}, which implies a
stellar mass of $2.5\times10^{8}$ \msun. While the best fit, as defined by the $\chi^2$ parameter, is useful for visualisation of the model spectrum it does not
represent the range of acceptable solutions or provide an uncertainty in the process. Therefore {\sc MAGPHYS} also calculates the probability density function
over a range of model values and determines the median and 
confidence interval corresponding to the 16th-84th percentile range. This is equivalent to the 1$\sigma$ range, if the distribution is roughly gaussian \citep[see][for details]{2008MNRAS.388.1595D}. As recommended
by \citet{2008MNRAS.388.1595D}, we take the best estimate of stellar mass to 
be this median $2.8 \times 10^{8}$ \msun and the 1$\sigma$
range to be from $2.3\times 10^{8}$ to $3.7 \times 10^{8}$ \msun.

We examined the effect of our correction for emission line contamination and how the wavelength range of input photometry influences the results of {\sc MAGPHYS} to determine the stellar mass of the galaxy. We used the {\it gri} photometry with and without the emission 
line contribution and we used different sets of wavelength restricted photometry as listed in Table\,\ref{tab:stellar_mass}. The results from using {\it ugriz} only and including the UV and NIR contributions are virtually identical which simply reflects the fact that the {\it ugriz} coverage is enough to constrain the SED and the UV, optical and NIR flux are internally consistent with typical galaxy SEDs. The major difference arises in the inclusion of the nebular emission, which results in typically a factor of $\sim$  2 difference in the median fits for mass. 
The $\chi^{2}$ values for the best fits are also significantly better
when using {\it gri} mags corrected for emission line flux. 
This illustrates the need to remove the emission line contribution and that this removal is more important than having wide wavelength coverage in the NUV and NIR. For comparison, we also fitted the galaxy SED model using the code {\sc Z-PEG} \citep{2002A&A...386..446L}. 
We again used the UV and optical photometry of the galaxy with the emission lines removed and found the stellar mass of $3.3\times10^{8}$ \msun\ for the best-fit.
The stellar mass estimation from {\sc Z-PEG} is consistent to within 10\% with that from the {\sc MAGPHYS}.

\begin{table}
\caption{Stellar mass determination from {\sc MAGPHYS}. We listed the different input photometry settings and the corresponding SED best-fit results. The main effect is removing or keeping the strong emission line contributions, which are equivalent to {\it g}=0.16, {\it r}=0.17 and {\it i}=0.71 magnitudes. }
\label{tab:stellar_mass}
\begin{tabular}[c]{llll}
\hline
 Input photometry & Emission  & Best-fit  & Median mass \\
 & line &$\chi^{2}$ & \msun  \\
\hline  
{\it FUV, NUV, ugriz,  JHK$_{s}$}       & excluded & 1.300 & $2.8\times10^{8}$  \\ 
                                       & included     & 3.428 & $4.4\times10^{8}$  \\ \cline{1-4}
{\it FUV, NUV, ugriz} & excluded & 1.258 & $2.9\times10^{8}$ \\ 
                        & included     &  3.640 & $7.0\times10^{8}$ \\ \cline{1-4}
{\it ugriz}                & excluded & 1.204 & $3.0\times10^{8}$  \\ 
                        & included     & 4.727 & $8.0\times10^{8}$ \\ \cline{1-4}
{\it gri}                   & excluded & 1.207 & $3.7\times10^{8}$  \\ 
                        & included     & 0.691 & $1.7\times10^{9}$   \\ 
\hline 
\end{tabular}
\end{table}

In summary we take the best estimate of the galaxy stellar mass to be $2.8^{+0.9}_{-0.5}\times10^{8}$\,\msun\ from the median and equivalent 1$\sigma$ uncertainty range from the probability density function of {\sc MAGPHYS}. 

For comparison we carried out a similar {\sc MAGPHYS} analysis for the host galaxy of SN~2007bi with the observed {\it griz} photometry as discussed in section\,\ref{sec:host_galaxy_photometry}. 
The stellar mass from the best-fit model (which has a $\chi^{2}=0.031$) is $1.2\times10^{8}$\,\msun with $z = 0.127$. 
The $1\sigma$ likelihood distribution was from $1.1\times10^{8}$ to $2.0\times10^{8}$ around the median value of $1.4\times10^{8}$\,\msun, which we adopt in the rest of this paper.

\subsection{Age of the stellar population}
\label{sec:stellar_population}

{\sc MAGPHYS} also calculates the 
{\it r}-band light-weighted age of the host stellar population. This is effectively an average age over all the stars in the galaxy. 
The light-weighted age is quantitatively defined in the equation from \citet{2005MNRAS.362...41G}: 
\begin{equation}
t_{r}= \int_{0}^{t} [d\tau\psi(t-\tau) f_{r}(\tau)\tau] \, / \int_{0}^{t}[d\tau\psi(t-\tau)f_{r}(\tau)] 
\end{equation}
where $f_{r}(\tau)$ is the total {\it r}-band flux radiated by stars of age $\tau$, and $\psi(t)$ is the star-formation rate.
{\sc MAGPHYS} gives an age of 44.7 Myr for the host of PTF12dam.
For comparison, we also obtain stellar population age estimates using two other common methods.

Firstly we used the \citet{2005MNRAS.362..799M} simple stellar population (SSP) models, which determine the age of the dominant stellar population in the host. We again used the synthetic photometry described above, after foreground extinction correction and removal of the host galaxy emission lines and SN continuum. 
We employed a metallicity of 0.5 \zsol\, and assumed a Salpeter IMF with red horizontal branch morphology. The best-fitting result for the stellar population age is between 10 and 30 Myr. 

Secondly, we used the rest-frame EW of \ha ($EW_{rest} = 527$ \AA) to compare with the $starburst99$ models \citep{1999ApJS..123....3L}, assuming an instantaneous burst of star formation with a Salpeter IMF at solar metallicity. Such methods have also been employed recently to analyse the spectra of the environments of local supernovae \citep{2013AJ....146...30K}. This resulted in a stellar population age around 30 to 40 Myr for the host of PTF12dam. 

Taking these three estimates together, there appears to be a 
a dominant young stellar population in the host, which is likely a common trend for SLSN Ic hosts. For instance, the stellar age for the host of 
SN~2010gx was estimated to be 20-30 Myr \citep{2013ApJ...763L..28C}, and 5 Myr for the host of PS1-10bzj \citep{2013ApJ...771...97L}. 
This is not unexpected as the working assumption is these SLSNe Ic arise from massive stars. 
However a quantitative restriction on the progenitor age (or mass) is not possible. The large range in the age estimates of
the underlying population, and the fact that multiple epochs of star formation could have 
occurred in the last few Myrs, means that more restrictive information on the progenitor systems is unlikely to be forthcoming
from these population age studies. All that can confidently be 
determined is that a large population of stars younger than 30-40\,Myr
is present in the host of PTF12dam. 

\subsection{Metallicity of the host galaxy}
\label{sec:metallicity}

We carried out an abundance analysis of the host galaxy based on the determination of the electron temperature derived from the \Oiii $\lambda4363$ auroral line. 
Firstly, the observed emission line flux (Table\,\ref{tab:emission_line_flux}) were corrected for the foreground galactic ({\it A$_{V}$} = 0.03) and internal dust ({\it A$_{V}$} = 0.20) extinctions.

The \Oiii $\lambda4363$ auroral line in the WHT spectrum (Fig.\,\ref{fig:12dam_host_spec}) is a strong, high signal-to-noise detection allowing us to estimate the electron temperature directly. 
We employed the task {\it temden} within the {\sc iraf/stsdas} package to calculate the electron temperature from the line ratio of $\lambda\lambda4959,5007$/$\lambda4363$, given the electron density $n_{\rm e}=106$\,cm$^{-3}$ from the line ratio of \Sii 
$\lambda\lambda 6717/6731=1.31$. We thus determined $T_{\rm e}$(\Oiii) $\sim 13500$\,K$ \equiv T_{3}$. 
Assuming a two-zone representation of the ionised region, this electron temperature is taken as the temperature in the high ionisation zone ($T_{3}$), and the value for electron temperature in the low ionisation zone ($T_{2}$) is determined as follows \citep[from][]{1992AJ....103.1330G}: 
\begin{equation}
 \label{eq:t2_t3_relation}
  T_{2} = 0.7T_{3} + 3000 
\end{equation}
We thus obtained an $T_{2} \sim 12400$\,K.

As we have the electron temperature, electron density and flux ratios of \Oii $\lambda3727$, \Oiii $\lambda5007$ lines relative to \hb respectively, we determined the ionic abundance of O$^{+}$ and O$^{++}$ using the {\it nebular.ionic} routine within the {\sc iraf/stsdas} package. 
The total oxygen abundances were then derived from the following equation: 
\begin{equation}
  \frac{O}{H} = \frac{O^{+}}{H^{+}} + \frac{O^{++}}{H^{+}}
\label{eq:oxygen_abundances}
\end{equation}
This provides an oxygen abundance of $12 + \log{\rm (O/H)} = 8.04 \pm 0.09$ for the host of PTF12dam.
For a solar value of $8.69 \pm 0.05$ \citep{2009ARA&A..47..481A}, this would imply that the host galaxy is 0.67 dex below solar abundance or 0.2 \zsol.

Due to the detection of \Heii $\lambda4686$, we checked if the 
amount of oxygen in the form of O$^{+++}$ would influence the total oxygen abundance. 
The ionization potential of O$^{++}$ is similar to that of He$^{+}$ 
\citep[54 eV;][]{1989agna.book.....O, 2006agna.book.....O}, hence we can simply
estimate the amount of O$^{+++}$ from the ratio of the \Heii and \Hei recombination
lines. We find that this would only increase the total oxygen abundance 
by 0.01 dex compared to the value from equation \ref{eq:oxygen_abundances}, and hence is a 
negligible contribution.

We also checked this with an alternate method for estimating $T_{2}$. 
Instead of calculating $T_{2}$ indirectly from equation\,\ref{eq:t2_t3_relation}, we used the ratio of nebular lines \Oii $\lambda\lambda3726,3729$ to auroral lines \Oii $\lambda\lambda7320, 7330$ to estimate the electron temperature $T_{2} \sim 14100$\,K directly, which was derived from the {\it temden} task. 

This method gives a value for $T_{2}$ which is somewhat higher 
than that for $T_{3}$, and strictly is not a consistent physical solution. However it is not
uncommon in large observed galaxy samples \citep[e.g. Fig.\,4a in ][]{2006A&A...448..955I} to see significant scatter in the $T_{2}$ and $T_{3}$ correlations and to have this situation. The reasons are 
unclear since the measurement uncertainties in the flux of 
all the lines are not too large ($\sim$10\% at most). 
Despite this concern, if we calculate the abundance of O$^{+}$ and O$^{++}$ using $T_{2} \sim 14100$\,K and $T_{3} \sim 13500$\,K, we determined a value of $12 + \log{\rm (O/H)} = 8.00$, which is within the uncertainty of the value obtained initially.

As the nebular and auroral \Siii lines in the NIR and optical $\lambda\lambda9069,9532$/$\lambda6312$ were detected, we checked the electron temperature estimate from these for consistency. This gave an approximate value of $\sim 16400$\,K, which is significantly higher than that derived from \Oiii lines. We believe the discrepancy is due to the lower signal-to-noise of the \Siii lines, the fact that there is significant SN continuum in the NIR spectrum (since this spectrum was taken on +27d after the peak), and the flux calibration is not as reliable as for our optical spectrum.

In order to calculate the sulphur abundance, we adopted the electron temperature derived from the \Oiii lines ($T_{3} \sim 13500$\,K) and applied equation\,\ref{eq:t2_t3_relation} to obtain $T_{2} \sim 12400$\,K. We used the {\it ionic} task to determine the $S^{++}/H^{+}$ and $S^{+}/H^{+}$ from the flux ratio of \Siii $\lambda6312$ and \Sii $\lambda6717$ lines relative to \hb. We assumed that the total sulphur abundance is the sum of these two dominant ionic species, to give $12 + \log{\rm (S/H)} = 6.17 \pm 0.12$. Hence, we calculated the $\log{\rm (S/O)} = -1.87 \pm 0.12$ (using the $T_{\rm e}$ based oxygen abundance). This is slightly lower than the typical ratios in metal-poor galaxies in \citet{2006A&A...448..955I}, which show are in the range $\log{\rm (S/O)} = -1.7 \pm 0.2$.

There are not many SLSN Ic host galaxies with metallicities determined by the direct method with electron temperature measurements. Hence
in order to make comparisons, we also determined the metallicities from several commonly used diagnostics. The application of the $R_{23}$ (M91) strong line method with the \citet{1999ApJ...514..544K} calibration, provides two possible estimates depending on whether the galaxy sits on the upper or lower branch of the calibration curve. 
These are 8.46 and 8.18 respectively. Since we have detected the auroral line, we can assume that we should be on the lower metallicity branch of the M91 relation. Hence the $R_{23}$ oxygen abundance estimate is $12 + \log{\rm (O/H)} = 8.18 \pm 0.03$. 
The other commonly used metallicity diagnostic is the N2 method (\citealp{2004MNRAS.348L..59P}, hereafter PP04) which uses the N2 index = $\log$(\Nii$\lambda6583$/\ha). This method has the advantage that it is less affected by dust extinction, given the close wavelengths of the emission lines. The N2 index has a positive correlation with oxygen abundance and the metallicity derived from the PP04 calibration is consistent with that from the $T_{\rm e}$ method \citep{2007A&A...473..411L}; we determined $12 + \log{\rm (O/H)} = 8.10 \pm 0.02$ with the N2 method.

We further obtained a nitrogen-to-oxygen ratio $\log{\rm (N/O)} = -1.32 \pm 0.13$ from the $T_{\rm e}$ calibration method, assuming that N/O = N$^{+}$/O$^{+}$. This places the galaxy close to the ``plateau'' seen at $\log{\rm (N/O)} \simeq -1.5$ on the N/O versus O/H diagram of galaxies \citep{2003A&A...397..487P}. For comparison, the mean values of oxygen and nitrogen abundances in 21 \Hii regions of the Small Magellanic Cloud are $8.07\pm0.07$ ($T_{\rm e}$ method) and $\log$(N/O) of $-1.55\pm0.08$. 
\citep[from literature values summarized in][]{2003A&A...397..487P}. Hence the metallicity environment in the host galaxy of PTF12dam appears quite similar to that in the Small Magellanic Cloud.

Using the SN as an illuminating source provides a way to study the ISM inside the host galaxy. For example, \Mgii $\lambda\lambda$2796/2803 have been detected in some SLSNe Ic hosts \citep{2011Natur.474..487Q}, and have been used to estimate the upper limit to the column density \citep{2012ApJ...755L..29B}. Recently, \citet{2014ApJ...797...24V} detected resolved \Mgii absorbers to study the host of iPTF13ajg. They found the strength of the \Mgi and \Mgii absorption in SLSN hosts is lower than in GRB hosts.
Here we detected absorption of the \Mgii $\lambda\lambda$2796/2803 doublet, which we presume is from the host galaxy. The two lines have the rest-frame EWs of $2.11\pm0.27$ and $2.01\pm0.28$ \AA\ respectively in the WHT spectrum on 2012 May 25 at the SN phase: $-13.6$d from the peak \citep{2013Natur.502..346N}. The \Mgii line centroids give almost identical redshifts ($z = 0.106$) to the \Oiii emission lines ($z = 0.107$) detected in the same spectrum. The ratio of their measured line strengths is W$_{2796}$/W$_{2803}$ = 1.05. 
The ratio of the oscillator strengths of these two transitions is close to 2, hence they are likely to be on the saturated part of the curve of growth which limits their use in diagnosing column densities in the host.

\begin{figure*}
       \centering
       \begin{subfigure}
           \centering
           \includegraphics[angle=0,width=0.48\textwidth]{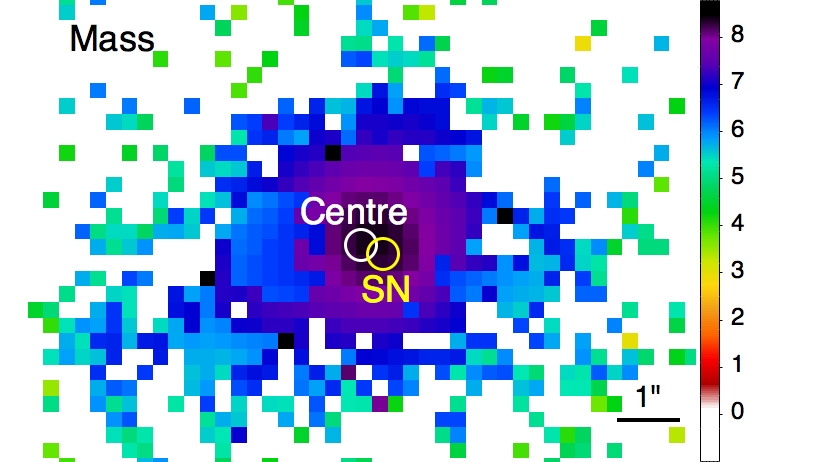}
       \end{subfigure}
       \begin{subfigure}
           \centering
           \includegraphics[angle=0,width=0.48\textwidth]{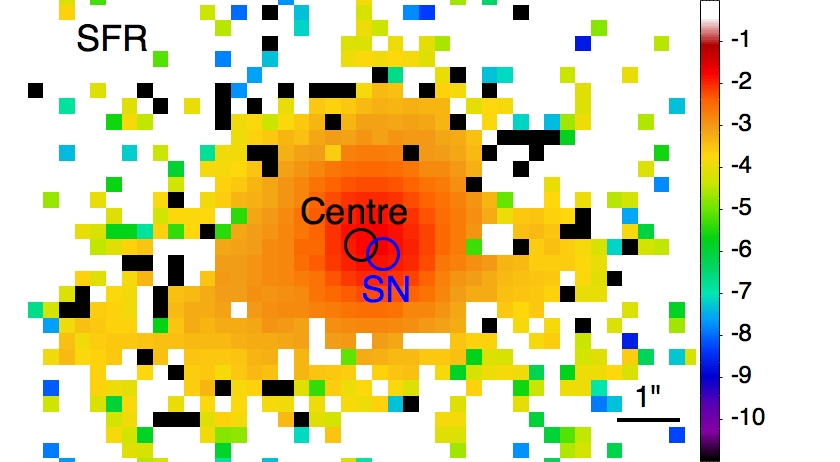}
       \end{subfigure}
\caption{Five-band ({\it ugriz}) pixel-pixel study of PTF12dam host galaxy. {\it Left:} The variation in stellar mass distribution of the host. The colour bars indicate stellar mass from $10^{0}$ to $10^{8}$ \msun. {\it Right:} The variation in star-formation rate distribution of the host. The colours show levels of star-formation rate from $10^{-1}$ to $10^{-10}$ \msun\,yr$^{-1}$. The host centre is defined by the brightest region of the combined optical band image and marked with a 0.25\arcsec size circle, the SN position is $ \sim 0.4 \pm 0.5$\arcsec around the centre. The centre and SN positions are located in the densest areas of the stellar mass and SFR maps.}
\label{fig:pixel_study}
\end{figure*}

\subsection{Star-formation rate}
\label{star_formation_rate}

After Milky Way and internal extinction correction, the observed 
flux of \ha is $F_{H\alpha} = 2.29\times10^{-14}$ $\rm{erg\,s^{-1}\,cm^{-2}}$. Applying the luminosity distance of $d_{\rm L} = 481.1$ Mpc \citep{2006PASP..118.1711W} of the host redshift ($z = 0.107$), we determined the luminosity $L_{H\alpha} = 6.35\times10^{41}$ $\rm{erg\,s^{-1}}$. 
We employed the calibration of \citet{1998ARA&A..36..189K} : 
\begin{equation}
 \rm{SFR} (\msun\,\rm{yr}^{-1}) = 7.9\times10^{-42}\,L(\ha) (\rm{erg\,s^{-1}})
\end{equation}
to determine the star-formation rate of the galaxy to be 5.0 \msun$\,\rm{yr}^{-1}$. We also estimated the SFR from the \Oii $\lambda3727$ line flux for comparison, again using a calibration from \citet{1998ARA&A..36..189K} :

\begin{equation}
 \rm{SFR} (\msun\,yr^{-1}) = (1.4\pm0.4)\times10^{-41}\,L(\Oii) (\rm{erg}\,s^{-1})
\end{equation}

After correction for Milky Way and internal extinction and luminosity distance, the luminosity of the \Oii line is $L_{\rm [OII]} = 3.68\times10^{41}$ $\rm{erg}\,s^{-1}$ which gives $5.2\pm1.5$ \msun$\,yr^{-1}$. This is consistent with the \ha measurement which further supports the use of \Oii $\lambda$3727 as a SFR diagnostic in higher redshift galaxies when H$\alpha$ may be inaccessible if it is shifted into the NIR regime \citep{2012ApJ...755L..29B}. 

The GALEX {\it FUV} flux provides another way to estimate the SFR and to check for consistency between the methods
\citep[e.g. see][for a discussion of the comparison between the methods in the local volume]{2012A&A...537A.132B}. Corrections were applied for 
Milky Way extinction ({\it A$_{FUV}$} = 0.09, {\it A$_{NUV}$} = 0.10), {\it K}-correction ($-0.05$ mag for {\it FUV}, 0 mag for {\it NUV}), and internal extinction ({\it A$_{FUV}$} = 0.39, {\it A$_{NUV}$} = 0.46, assuming {\it R$_{V}$} = 4.05) using the algorithm determined by \citet{1989ApJ...345..245C} since \citet{2011ApJ...737..103S} do not provide UV wavelength calibrations. 
We calculated a luminosity of L$_{1380 \AA} = 1.32 \pm 0.26 \times10^{28}$ ($\rm{erg}\,s^{-1}\,Hz^{-1}$), and then used the conversion of \citet{1998ARA&A..36..189K} to estimate the star-formation rate:
\begin{equation}
 \rm{SFR} (\msun\,yr^{-1}) = 1.4\times10^{-28}\,\rm{L_{FUV}} (\rm{erg}\,s^{-1}\,Hz^{-1})
\end{equation}
which gives SFR $= 1.8 \pm 0.4$ (\msun$\,yr^{-1}$). This is somewhat lower than that derived from H$\alpha$ (5.0 \msun$\,yr^{-1}$), and the discrepancy is in the opposite sense to that found by the \cite{2012A&A...537A.132B} study which concluded that SFRs based on $L_{\rm FUV}$ appeared to be systematically higher (by nearly a factor 2) than those from $L_{\rm H\alpha}$.

Since the FUV luminosity is quite significantly affected by the internal dust extinction law (and value) applied, we considered that the \ha luminosity is less affected by dust extinction and there are more dwarf galaxy measurements to compare this method with, we will adopt the SFR = 5.0\,\msun$\,\rm{yr}^{-1}$ from \ha method as the value for the PTF12dam host galaxy throughout the rest of this thesis. 
Finally, the specific star-formation rate (sSFR) is defined as the star-formation rate divided by stellar mass. We calculated the sSFR of the PTF12dam host to be 17.9 Gyr$^{-1}$.

By comparison, the SFR in the host of SN~2007bi is only 0.01 \msun$\,\rm{yr}^{-1}$ from the \ha flux reported in \citet{2010A&A...512A..70Y}; the PS1-11ap host has
SFR = $0.47\pm0.12$ \msun$\,\rm{yr}^{-1}$ estimated from the \Oii$\lambda$3727 line flux.

\begin{table}
\centering
\begin{minipage}{83mm}
\caption{Main properties of the host galaxy of PTF12dam. All magnitudes listed are in the AB system, while the NIR magnitudes are converted from the Vega system \citep{2007AJ....133..734B}. Internal extinction with the rest frame wavelength of {\it g} band.}
\label{tab:property}
\begin{tabular}[t]{lllllllllllll}
\hline
SDSS name & SDSS J142446.21+461348.6 \\
\hline
RA (J2000) & 14:24:46.20 \\
Dec (J2000) & +46:13:48.3 \\
Redshift & 0.107 \\
GR6 GALEX {\it fuv} (mag) & $20.13\pm0.19$ \\
GR6 GALEX {\it nuv} (mag) & $20.13\pm0.14$ \\
DR9 Petro {\it u} (mag) & $19.79\pm0.05$ \\ 	
DR9 Petro {\it g} (mag) & $19.36\pm0.02$ \\ 	
DR9 Petro {\it r} (mag) & $19.20\pm0.02$ \\
DR9 Petro {\it i} (mag) & $18.83\pm0.03$ \\
DR9 Petro {\it z} (mag) & $19.32\pm0.14$ \\
{\it J} (mag) & $18.37\pm0.15$ \\
{\it H} (mag) & $17.82\pm0.13$ \\
{\it K$_{s}$} (mag) & $17.16\pm0.09$ \\
{\it K}-correction$_{g}$ (mag) & $\sim 0$ \\
Internal extinction$_{g}$ (mag) & $\sim 0.20$ ({\it R$_{V}$} = 4.05)\\
Luminosity distance (Mpc) & 481.1 \\
{\it M$_{fuv}$} (mag) & $-18.86\pm0.20$ \\
{\it M$_{nuv}$} (mag) & $-18.85\pm0.14$ \\
{\it M$_{u}$} (mag) & $-18.97\pm0.07$ \\
{\it M$_{g}$} (mag) & $-19.33\pm0.10$ \\
{\it M$_{r}$} (mag) & $-19.42\pm0.04$ \\
{\it M$_{i}$} (mag) & $-19.68\pm0.06$ \\
{\it M$_{z}$} (mag) & $-19.21\pm0.14$ \\
{\it M$_{J}$} (mag) & $-19.26\pm0.15$ \\
{\it M$_{H}$} (mag) & $-19.27\pm0.13$ \\
{\it M$_{K_{s}}$} (mag) & $-19.64\pm0.09$ \\
Physical diameter (kpc) & $\sim 1.9$ \\ 
$12+\log {\rm(O/H)}$ ($T_{\rm e}$) (dex) & $8.05\pm0.09$ \\
$12+\log {\rm(O/H)}$ ($R_{23}$) (dex) & $8.18\pm0.03$ \\
$12+\log {\rm(O/H)}$ (N2) (dex) & $8.10\pm0.02$ \\  
N/O ($T_{\rm e}$) & $-1.32\pm0.13$ \\
S/O ($T_{\rm e}$) & $-1.87\pm0.12$ \\
H$\alpha$ luminosity (erg s$^{-1}$) & $6.35\times10^{41}$ \\
SFR (M$_{\odot}$ year$^{-1}$) & 5.0 \\
Stellar mass (M$_{\odot}$) & $2.8\times10^{8}$ \\
sSFR (Gyr$^{-1}$) & 17.9 \\
\hline 
\end{tabular}
\end{minipage}
\end{table}

\subsection[]{Pixel by pixel analysis}
\label{sec:pixel_by_pixel_analysis}

Since the SLSNe Ic in \cite{2014ApJ...787..138L} show a wide diversity in the positions within the {\it Hubble Space Telescope (HST)} images 
of some of the host galaxies, we undertook a detailed pixel by pixel analysis of the host of PTF12dam to determine if it
was located in the highest star-forming and densest stellar mass regions. 

We used the optical {\it ugriz}-band images taken from the WHT + ACAM on 2014 January 22, +534d (rest-frame) after the peak. We assumed that the SN flux is negligible in the images. To extend the wavelength coverage for the SED fitting, we also took {\it u} and {\it z}-band images (specifically the WHT ACAM filters: 700 SlnU, 704 SlnZ). 
We adopted 13 reference stars from the SDSS DR9 catalog \citep{2012ApJS..203...21A} to set the photometric zero points.
Instead of measuring the whole galaxy flux, 
we chose an area ($50\times50$ pixels) around the host galaxy centre and calculated their flux values pixel by pixel for each {\it ugriz}-band images. The stellar mass and SFR of each pixel were determined using {\sc MAGPHYS} 
and we treated each individual pixel as a ``galaxy'' in the {\sc MAGPHYS} SED fitting: 
Five-band ({\it ugriz}) flux of each pixel were compiled and fed into the {\sc MAGPHYS} software. 
The input flux were Milky Way extinction corrected according to \citet{1998ApJ...500..525S}. 
There were no internal extinction priors assumed in the SED fitting. 
Note that since the emission line distribution in the host is not resolved, the flux are not emission line free.
 
Marginalised likelihood distribution of each physical parameter, e.g. stellar mass and SFR, of that pixel are then determined by comparison of the input SED with all the model galaxies' SEDs in the {\sc MAGPHYS} library. Both SFR and stellar mass are determined using SED fitting based on the photometry of all five bands in {\sc MAGPHYS}, see section\,\ref{sec:stellar_mass}.

The maps are shown in Fig.\,\ref{fig:pixel_study} which give the median value (50\%) of the likelihood distribution, not the value of the best-fit model. 
The SN position is very close to the optical centre of the host with a small $0.4\pm0.5$\arcsec offset. 
The most dense region in the stellar mass map is close to the optical centre of the host (as is the SN 
position) at a value corresponding to $10^{8.4}$\msun. This also is the region of highest SFR ($0.02$ \msun$\,yr^{-1}$). 
Recently, \citet{2015ApJ...804...90L} presented pixel statistics for 16 SLSN Ic hosts using {\it HST} UV images. They found that SLSNe Ic occur in overdense regions, in which a median SFR density is $\sim 0.1$ \msun\,yr$^{-1}$\,kpc$^{-2}$. However, in our case, the ground-based host images are seeing dominated, which means one should not interpret the real galaxy scale as being spatially resolved at finer detail in these images.

\section[]{The bolometric lightcurve of PTF12dam from $-$52 to +399 days}
\label{bolometric_light_curve_of_12dam}

\cite{2013Natur.502..346N} presented a bolometric lightcurve of PTF12dam 
from $-52$d to +222d which included contributions from the 
UV ($\lambda_{c} = 1928-2600$ \AA), optical ({\it ugriz}) and NIR ({\it JHK$_{s}$}). 
Here we extend this baseline out to 400 days (rest-frame from the peak), to further test the competing models that have 
been developed to explain the extreme luminosity. 

We used our measured late epoch {\it gri} and {\it JHK$_{s}$} magnitudes (listed in Table\,\ref{tab:phot}) to 
calculate the bolometric luminosity of PTF12dam at +200-400d in the same way as \citet{2013Natur.502..346N}. 
The magnitudes were adjusted for Milky Way extinction ({\it A$_{V}$}=0.04), {\it K}-corrections 
\citep[calculated from the +221d spectrum of][]{2013Natur.502..346N} and internal dust extinction of the host ({\it A$_{V}$}=0.2). We chose to calculate the bolometric flux at every epoch with an {\it r}-band 
observed data point\footnote{For the final 
epoch +399d, we had only an upper limit in the {\it r} band, and hence we used the {\it i}-band magnitude as the primary reference magnitude instead.}. At these epochs, magnitudes in the available filters were converted to physical flux, and an SED was then constructed by linearly interpolating between the observed flux. 
If any of the {\it giJHK$_{s}$} filter measurements were not available at the defined epoch, we estimated the 
magnitude by linear interpolation between the two closest points in time. This method provides a reliable measurement of 
bolometric flux from the {\it g} band through to {\it K$_{s}$}, but does not account for flux beyond these limits. 
Since the lightcurve models of magnetar powering, pair-instability or CSM interaction provide calculations
of total bolometric flux output \citep{2010ApJ...717..245K,2011ApJ...734..102K,2013Natur.502..346N, 2014MNRAS.444.2096N} we require a method to account for the UV and mid infra-red (MIR) contributions. 
The methods to determine the full bolometric lightcurve between 1600 \AA\ to 4.5$\mu$m are described below.
Our analysis principle is to first use real, measured flux where possible. When data points are missing, we try to 
use measured data from similar SNe, and finally if that fails we use blackbody fits to cover portions of the SED where
no observational data exists.

PTF12dam was observed by the Ultraviolet and Optical Telescope (UVOT) on the Swift and ground based {\it u} band covering $\sim$ 1600-3800 \AA\ but only until +22d. To determine the UV contribution after this, we have a ready made template in the supernova PS1-11ap. This SN is at $z = 0.524$ and the observer frame {\it g$_{\rm P1}$} and {\it r$_{\rm P1}$} Pan-STARRS1 filters correspond approximately to the rest-frame filters {\it UVW1$_{\rm Swift}$} and {\it u$_{\rm SDSS}$} \citep[see][]{2014MNRAS.437..656M}. This provides measured {\it UVW1$_{\rm Swift}$} data to +83 days and 
{\it u$_{\rm SDSS}$} data to +254d. We assume that the relative UV flux of PTF12dam (compared to the optical)
is similar to that of PS1-11ap which is justified given their similar spectra, lightcurve and colour evolution
\citep{2013Natur.502..346N,2014MNRAS.437..656M}. 
This provides measured flux for PS1-11ap down to 2250 \AA\ in the near-UV until +83d and
to down to 3000 \AA\ until +254d. To calculate the flux contribution down to 1600 \AA\ we fitted blackbody
spectra to the observed SEDs defined by these filters ({\it UVW1,ugri}). This provided 
the flux contribution in the UV range (1600-3100 \AA) compared to the optical of PS1-11ap during $-36$d to +254d.
In summary, 
we used the measured Swift UV data for PTF12dam when available (until +22d) and used the UV flux 
contribution estimated for PS1-11ap from +25d to +254d. 
Using the values between +100 and +254d, we linearly extrapolated the UV contribution to provide estimates between
+245d and +399d. Hence we have a reasonable estimate of the UV flux of PTF12dam 
across the whole lightcurve of $-52$ to +399d. The relative contribution is plotted in Fig.\ref{fig:12dam_fraction}, 
showing the region of linear extrapolation. We know that the 
UV contribution to PTF12dam is not insignificant at these
late times since the observed spectrum slope is 
of PTF12dam at +221d is still rising and has not turned over at 3000 \AA\ \citep[Fig.\,2 in][]{2013Natur.502..346N}. 

Further corrections are required to properly account for flux in the NIR and MIR. PTF12dam was 
observed in the NIR {\it JHK$_{s}$} bands to +258d, allowing a full integration of the observed flux out to 2.3 $\mu$m
over this period. A {\it J}-band measurement was available until +342d, and an {\it H}-band measurement until +289d. We
used these points, but otherwise the {\it JHK$_{s}$} bands were linearly extrapolated to the dates necessary to allow
a contribution to be calculated. Observations of the MIR flux of supernovae have been limited to the very nearby explosions. 
We assume that the MIR contribution (i.e. beyond the red-edge of the {\it K$_{s}$} filter) for PTF12dam is negligible up to 150d, but that it becomes significant afterwards which is supported by observations of 
nearby SNe (e.g. SN~2011dh, \citealt{2014A&A...562A..17E}; SN~1987A, \citealt{1989A&AS...80..379B}). 
 For all epochs after +150d from the peak, we simply assumed
that the MIR flux was 20\% of the total bolometric luminosity. 
 This 20\% estimation is based on the MIR contribution to type IIb SN~2011dh \citep{2014A&A...562A..17E} and is in reasonable agreement with the MIR contribution for SN~1987A in the early nebular phase \citep{1988AJ.....95...63H, 1989A&AS...80..379B}. The lack of MIR data for other hydrogen poor type Ibc SNe in the local Universe prevents any further comparison. We adopted the data from the type IIb SN~2011dh, since it is a stripped envelope object with a very low mass hydrogen envelope. 
If we assume that the SN at late-times was a blackbody emitter, the MIR flux contribution would be $\sim 20\%$ for a temperature of 3000 K, and $\sim 15\%$ for a temperature of 3500 K. Adoption of a 20\% contribution is admittedly somewhat arbitrary, but is probably the best estimate we have at this point. It is only likely to be significantly different if SLSNe Ic somehow show peculiar MIR emission compared to normal SNe. 

\begin{figure}
\includegraphics[angle=0,width=\linewidth]{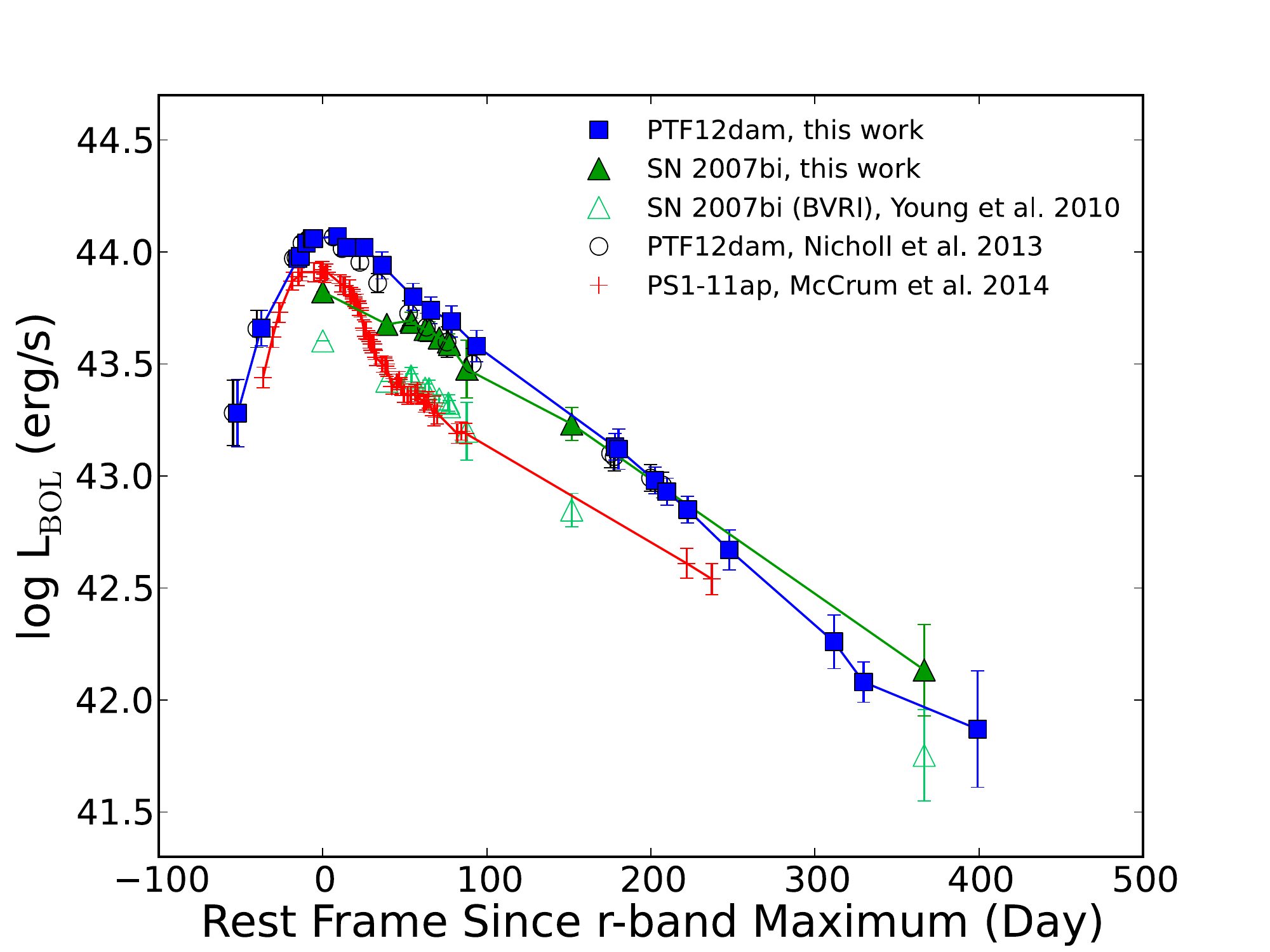}
\caption{Bolometric lightcurves of PTF12dam, SN~2007bi and PS1-11ap. A true bolometric lightcurve of 
PTF12dam (integrating the estimated SED from the FUV to MIR) is shown as the solid blue squares. 
For reference, the previously published calculations of \citet{2013Natur.502..346N} are plotted and
the differences are discussed in section\,\ref{bolometric_light_curve_of_12dam}. 
A similar, true bolometric lightcurve of SN~2007bi is shown using the $BVRI$ broad band photometry from \citet{2010A&A...512A..70Y} and a similar SED evolution as for PTF12dam. For comparison the 
pseudo-bolometric lightcurve presented in \citet{2010A&A...512A..70Y} is plotted, indicating what the 
the likely correction is if flux in the UV and NIR/MIR is ignored. 
The bolometric lightcurve of PS1-11ap as calculated in \citet{2014MNRAS.437..656M} is plotted as red crosses. 
This is directly comparable with the true bolometric lightcurves of PTF12dam and SN~2007bi, although
beyond +100-150d it may be underestimated by 10-20\% due to lack of inclusion of a MIR contribution.
 }
\label{fig:12dam_bol_lc}
\end{figure}

\begin{figure}
\includegraphics[angle=0,width=\linewidth]{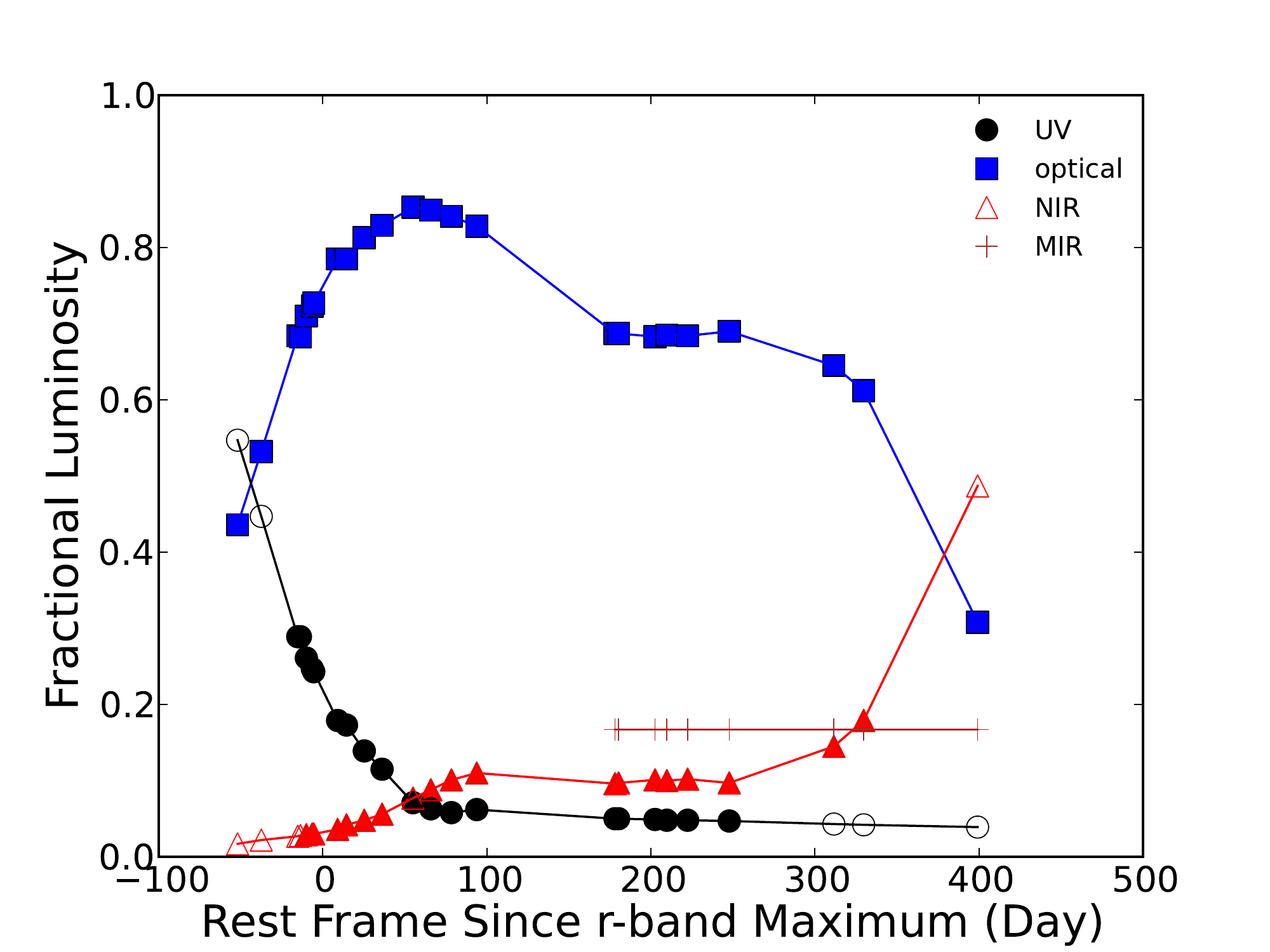}
\caption{Relative contributions of the UV, optical, NIR and MIR flux of PTF12dam to its total luminosity. The solid symbols indicate where measured data points were used and the hollow symbols show the flux which were based on the extrapolation methods described. The MIR contribution of 20\% of the total flux was assumed.}
\label{fig:12dam_fraction}
\end{figure}

The final bolometric lightcurve of PTF12dam is plotted in Fig.\,\ref{fig:12dam_bol_lc} 
and the data points are listed for reference in Table\,\ref{tab:bol_flux}. 
The fraction of total luminosity in the UV, optical, NIR and MIR wavelength ranges using this method
is illustrated in 
Fig.\,\ref{fig:12dam_fraction}. 
Uncertainties are accounted for in the error bars in Table\,\ref{tab:bol_flux}. These 
include the error in photometric measurements and an uncertainty arising from the extrapolation method for the missing filter data. The latter
was estimated by comparing two different methods. The first was when we used linear extrapolation (or interpolation) of observed lightcurve points to estimate a magnitude. The 
second was to assume a constant colour evolution from the nearest measured points in time (with respect to the {\it r} band) and apply this colour to get the missing filter data. 
The difference between the two is used as an error estimate.

\begin{table}
 \label{tab:bol_flux}
 \centering
 \begin{minipage}{83mm}
  \caption{Bolometric flux and uncertainty of PTF12dam and SN~2007bi. The SN phase is repored with respect to the observed maximum light in the rest-frame.}
\begin{tabular}{c c | c c}
\hline
PTF12dam &  & SN~2007bi & \\ 
\hline
Phase & $\log$ $L_{\rm Bol}$ & Phase & $\log$ $L_{\rm Bol}$ \\  
(day) & (erg/s) & (day) & (erg/s) \\  
\hline
-51.94&$43.28\pm0.15$&0.00	&$43.82\pm0.00$ \\
-37.49&$43.66\pm0.08$&39.04&$43.68\pm0.00$ \\ 
-15.24&$43.97\pm0.03$&53.77&$43.69\pm0.04$ \\
-13.62&$43.98\pm0.03$&54.13&$43.67\pm0.02$ \\
-9.98&$44.04\pm0.03$&62.38	&$43.65\pm0.02$ \\
-6.38&$44.06\pm0.03$&64.77	&$43.65\pm0.04$ \\
-5.47&$44.06\pm0.03$&70.98	&$43.62\pm0.02$ \\
9.06&$44.07\pm0.03$&76.57	&$43.60\pm0.04$ \\
14.47&$44.02\pm0.03$&77.20	&$43.59\pm0.03$ \\
25.30&$44.02\pm0.03$&87.76	&$43.48\pm0.13$ \\
36.16&$43.94\pm0.06$&151.73&$43.23\pm0.07$ \\
55.04&$43.80\pm0.06$&366.64&$42.13\pm0.20$ \\
65.87&$43.74\pm0.06$&&\\
78.46&$43.69\pm0.07$&&\\
93.82&$43.58\pm0.07$&&\\
178.15&$43.13\pm0.06$&&\\
180.33&$43.12\pm0.09$&&\\
202.48&$42.98\pm0.06$&&\\
209.77&$42.93\pm0.06$&&\\
222.38&$42.85\pm0.06$&&\\
247.75&$42.67\pm0.09$&&\\
311.59&$42.26\pm0.12$&&\\
329.68&$42.08\pm0.09$&&\\
399.19&$41.87\pm0.26$&&\\
\hline 
\end{tabular}
\end{minipage}
\end{table}

There are differences between the bolometric luminosities calculated in this work and \citet{2013Natur.502..346N} as illustrated in Fig.\,\ref{fig:12dam_bol_lc}. We added the UV contributions based on PS1-11ap estimate after +25d when there were no UV observations, whereas \citet{2013Natur.502..346N} assumed a linear decline 
by extrapolating a fit to the last few observed points. Also we now have NIR {\it JHK$_{s}$} template images for the NOTCam images and can subtract the contribution of the host, which is significant at +200d. 
The NIR data points in \citet{2013Natur.502..346N} have some host contamination and hence they are higher by a factor of $1.3-1.5$ compared to our values, after host subtraction. 
However, this is offset due to the fact that we estimate a higher UV contribution. Overall then, the combination
of these two factors results in a similar bolometric lightcurve as shown in \citet{2013Natur.502..346N} for the
epochs up to +220d. 

To allow a like-for-like comparison, we estimated the bolometric lightcurve of SN~2007bi over the same wavelength range as PTF12dam,
using as similar a method as possible. We adopted the $BVRI$ photometry from \citet{2010A&A...512A..70Y}, de-reddened it by the Milky Way extinction of A$_{V}=0.08$, performed {\it K}-corrections using values from \citet{2010A&A...512A..70Y}, and assumed 
internal dust extinction is negligible (as suggested from \ha and \hb line ratio). 
Due to the paucity of data around peak, we simply assumed a constant colour with {\it R} for the {\it B}, {\it V} and {\it I} lightcurves. 
We also assumed that SN~2007bi has a similar multi-colour photometric evolution as PTF12dam, and therefore that the percentage contributions in the UV and NIR, as a function of time, are the same (as in Fig.\,\ref{fig:12dam_fraction}). 
We assumed a 20\% MIR contribution to the data at epochs later than +150d after peak. 
These are also plotted in Fig.\,\ref{fig:12dam_bol_lc}, and as expected we find higher values than \citet{2010A&A...512A..70Y}. 
This is simply due to our accounting for flux outside the {\it BVRI} bands. 

\section{Discussion : three alternative lightcurve models}
\label{sec:lightcurve_models}

\subsection{Magnetar model}
\label{sec:magnetar_trapping}

The extension of the bolometric lightcurve to these quite late phases allows us to test the magnetar and $^{56}$Ni powering mechanisms as both make different predictions for the very late phases. These new points show that 
the late-time luminosity of PTF12dam declines more rapidly than the spin-down luminosity of the magnetar model 
which was used to fit the early-time data in \cite{2013Natur.502..346N}. However this model assumes 100\% efficient trapping of the magnetar energy in the ejecta. 
For a magnetar with parameters from that fit ($M_{\rm ej} = 16$ M$_{\odot}$, P$_{ms} = 2.6$ ms, B$_{14} = 1.14\times10^{14}$ G, and using $\kappa = 0.1$ cm$^2$g$^{-1}$) to be able to reproduce the observed 
behaviour at late epochs, escape of $\gamma$-ray radiation must be postulated. This escape must begin at around +200d post-explosion. 

The trapping and thermalization of the high energy radiation being emitted from the interface of the pulsar wind nebula and the supernova ejecta is a poorly understood process, with only \citet{2013MNRAS.432.3228K} and \citet{2014MNRAS.437..703M} having discussed the physical situation for the case of fast-spinning and highly magnetised pulsars. Assuming that the opacity for the high-energy radiation is dominated by pair production (high-energy gamma rays) the relevant opacity is of order $\kappa_{\gamma} = 0.01$ cm$^2$g$^{-1}$ \citep{1992hea..book.....L}. 

Fig.\,\ref{fig:bol_model} shows the best fitting magnetar model assuming a trapping with this opacity and the trapping function of \citet{1982ApJ...253..785A}. The model does a good fit of reproducing the full lightcurve. The loss of trapping begins around 100 days and is a factor $\sim$ 10 at 400 days. In this case the parameters of the best fit model are different to those in \citet{2013Natur.502..346N}. 
We required M$_{\rm ej} = 10.50$ \msun, P$_{ms} = 2.72$ ms, B$_{14} = 0.69\times10^{14}$ G for our new magnetar model. 
The opacity of the ejecta and the physical nature of the energy coupling 
with the putative pulsar wind is an area worthy of further, more detailed study. Here we merely show that the lightcurve decline rate at +200-400d can plausibly be explained if high-energy gamma rays are created at the base of the SN ejecta, as the opacity for such radiation (pair production) leads to partial escape after a few months.

\subsection{Pair-instability supernova model}
\label{sec:pisn}

\begin{figure}
\includegraphics[angle=0,width=\linewidth]{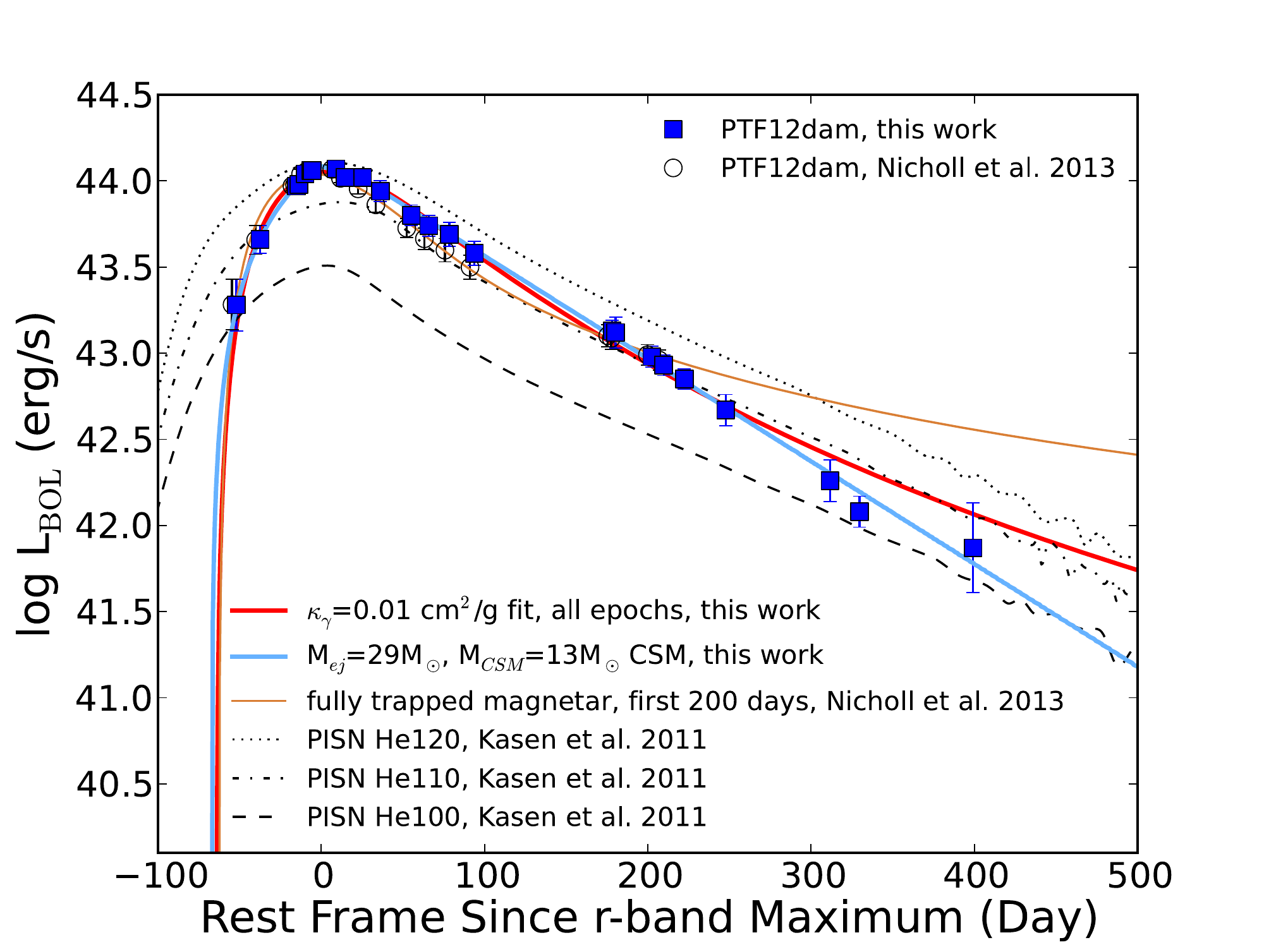}
\caption{Bolometric lightcurve of PTF12dam and alternative models. The new bolometric lightcurve of PTF12dam is shown with blue squares and the new magnetar best-fit model with $\kappa_{\gamma} = 0.01$ cm$^2$g$^{-1}$ is the red line. For comparison, the black circles show the bolometric lightcurve of PTF12dam from \citet{2013Natur.502..346N}, and their fully trapped magnetar model ($\kappa_{\gamma} = \infty$) is shown by the orange line. Background black lines illustrate PISN models \citep{2011ApJ...734..102K} of different mass, which are not fitted with the PTF12dam lightcurve evolution. Our CSM model (blue line) is also present. See section\,\ref{sec:lightcurve_models} for details. }
\label{fig:bol_model}
\end{figure}

In \citet{2013Natur.502..346N} it was shown that the rise time of PTF12dam is significantly shorter than in PISN models. 
Fig.\,\ref{fig:bol_model} shows that also the decline rate is faster than in PISN models \citep[e.g. ][]{2011ApJ...734..102K}. Due to their high ejecta masses, gamma ray trapping is complete in PISN models for several years, and their light curves therefore follow the decline rate of $^{56}$Co (1 mag per 100 days). PTF12dam on the other hand declines by over 1.5 mag / 100 days between 100 and 400 days. Thus, both the rising and declining parts of the light curve are inconsistent with a pair-instability ejecta, and can only be reproduced with models with significantly lower ejecta mass.

\subsection{CSM model}
\label{sec:csm}

An alternative model sees the light curve of PTF12dam powered by interaction of the SN ejecta with optically thick circumstellar gas. \citet{2014MNRAS.444.2096N} used the physical model of \citet{2012ApJ...746..121C} combined with the \citet{1982ApJ...253..785A} diffusion treatment to calculate a series of models powered by interaction. 
In this scenario, the collision of the rapidly expanding ejecta with a massive, slow-moving CSM (approximated as stationary) launches forward and reverse shocks into the CSM and ejecta, respectively, which convert the kinetic energy of the expanding ejecta to thermal energy behind the shocks. The observed light curve is then produced by outward diffusion of this energy. Shock heating terminates when the two shocks have passed through all of the CSM and ejecta, after which time (in the absence of additional power sources, such as $^{56}$Ni) the light curve decays exponentially. \citet{2014MNRAS.444.2096N} applied this model to the PTF12dam light curve up to 200 days after maximum, finding a good fit for M$_{\rm ej}$ = 26  \msun and M$_{\rm csm}$ = 13 \msun. 
Here we used the same model calculations as in \citet{2014MNRAS.444.2096N} and 
fitted the light curve up to day 400, with our improved UV, NIR and MIR corrections, and recover very similar fit parameters (M$_{\rm ej}$ = 29 \msun and M$_{\rm csm}$ = 13 \msun).

In order to distinguish the statistical significance of any difference between observations and models, we calculated the decline rate of the best-fit of our new bolometric data points of PTF12dam from +50d to +400d. Each data point was represented by Gaussian profile, centred on the measured luminosity and with a width of the error bars. We employed a Monte-Carlo simulation to add random points in these Gaussians, and fitted straight lines to these simulated points. This process was repeated 10000 times. Finally we measured the best-fit gradient of $0.015\pm0.001$ mag\,day$^{-1}$. The error includes both the dispersion in the points and the individual errors. 
We compared the best-fit result with the slope of PISN models, in the late-time phase, derived by $^{56}$Co decay of 0.0098 mag\,day$^{-1}$. The decline rate of these three PISN models is about 0.011 mag\,day$^{-1}$ around +100d to +400d, which is not consistent with the gradient of PTF12dam. For the fully-trapped magnetar model, the slopes are 0.011, 0.007 and 0.005 mag\,day$^{-1}$ during +100d to +200d, +200d to +300d and +300d to +400d, respectively. During the same duration, the slopes are 0.015, 0.012 and 0.010 mag\,day$^{-1}$ of our new magnetar model with $\kappa_{\gamma}$ = 0.01 cm$^{2}$\,g$^{-1}$. This fits with the data relatively well, and passes through all observed points, expected two points around +300d. The slope of CSM model is 0.015 mag\,day$^{-1}$, and this does match the best-fit gradient of PTF12dam. However, the CSM model required M$_{\rm ej}$ = 29 \msun and M$_{\rm csm}$ = 13 \msun. 
Such a large amount of CSM material has only been quantitatively predicted by the pulsational pair-instability SN (PPISN) model \citep{2007Natur.450..390W}, which requires a progenitor mass between 95 and 130 \msun \citep[also see ][]{2002ApJ...567..532H, 2003ApJ...591..288H}.

Overall, fitting of the full lightcurve is consistent with our new magnetar model with low $\kappa_{\gamma}$. This goes through the majority of our new bolometric lightcurve points and only two points were slightly discrepant. The ejecta mass estimated from the magnetar model is more than 10 \msun, hence the progenitor mass must be significantly larger. \citet{2003ApJ...591..288H} calculated that a massive single star more than 25 \msun should collapse to a black hole. This might suggest that the progenitor mass should be more than 10 \msun but less than 25 \msun to be consistent with the ejecta mass measurement, and stellar evolution theory.

\section{Discussion : the host galaxy of PTF12dam}
\label{sec:discussion}

\citet{2014arXiv1403.3441A} presented a large sample of 183 extreme emission-line galaxies (EELGs), which have large \Oiii $\lambda5007$ EWs and low-metallicities (median of $\sim$ 0.3 \zsol).
The host galaxy of PTF12dam would be defined as an EELG given the strengths of the classic nebular lines. 
For example the \Oiii $\lambda5007$ line is unusually strong and with EW$_{rest} \geq$ 563 \AA\ it has the highest 
EW of any host galaxy of SLSNe Ic within $z < 0.5$ 
\citep{2014ApJ...787..138L, 2015MNRAS.449..917L}. 
\citet{2015MNRAS.449..917L} measured the EW$_{rest}$ = 794 \AA\ for the \Oiii $\lambda5007$ line using their pure galaxy spectrum taken on +567d from the SN peak.
At {\it M$_{g}$} = $-19.33$ the host of PTF12dam is also the brightest 
host galaxy of any SLSNe Ic detected within $z < 0.5$ \citep{2014ApJ...787..138L, 2015MNRAS.449..917L}.

\begin{figure*}
 \includegraphics[angle=0,width=\linewidth]{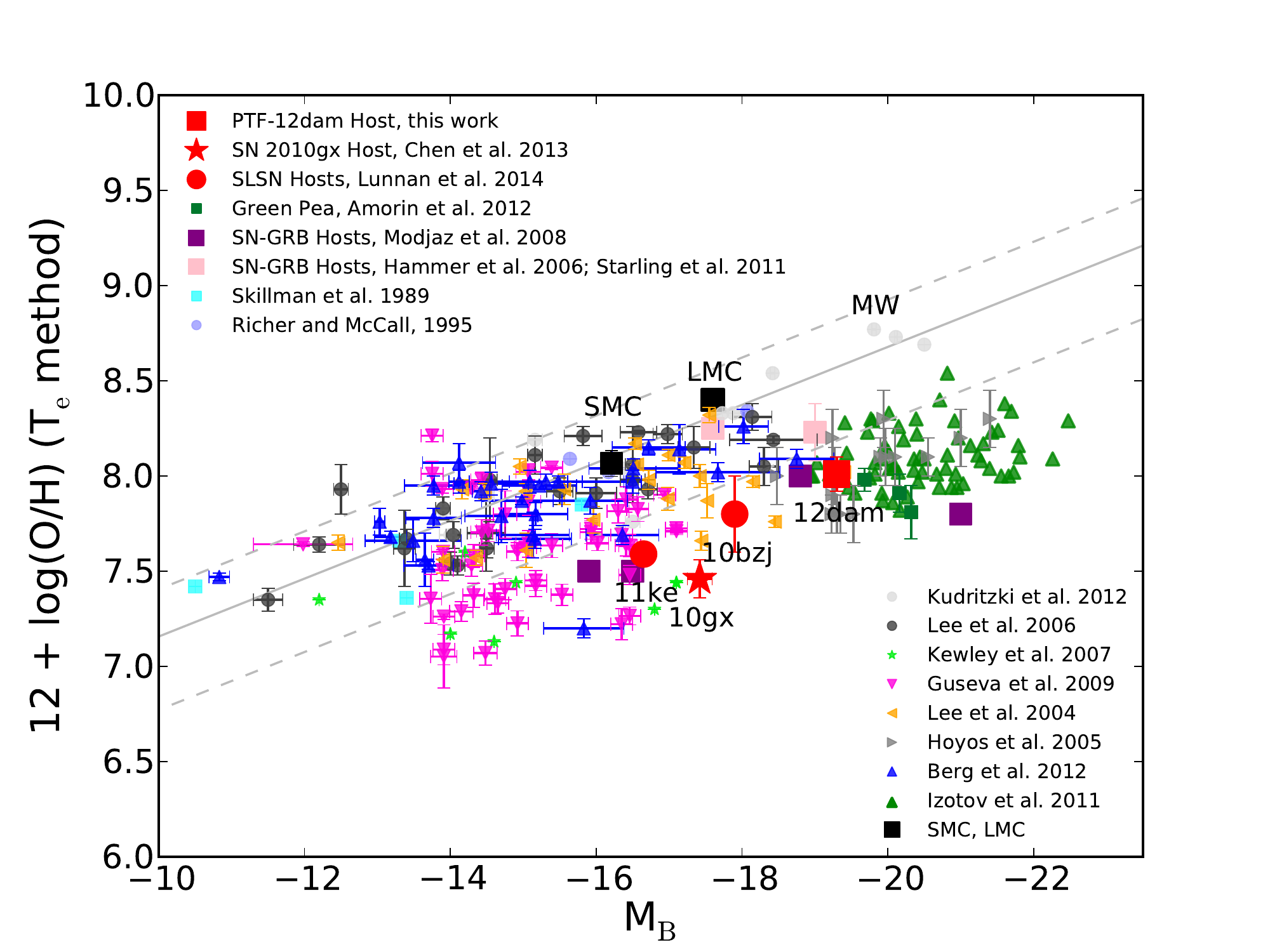}
 \caption{Luminosity-metallicity relationship for dwarf galaxies in the local Universe. Metallicity measurement was derived from the $T_{\rm e}$ direct method. Only four host galaxies of SLSNe Ic have direct oxygen abundance measurement so far; they all lie on the metal-poor edge compared with the local dwarf galaxies.
We adopted metallicity values from the literature as follows: \citealt{1989ApJ...347..875S, 1995ApJ...445..642R, 2004ApJ...616..752L, 2005ApJ...635L..21H, 2006ApJ...647..970L, 2007AJ....133..882K, 2009A&A...505...63G, 2012ApJ...749..185A, 2012ApJ...754...98B}. 
Gray lines are the boundary of normal galaxy mass-metallicity relation adapted from \citet{2012ApJ...747...15K}, with metallicities calculated using blue supergiants. GRB-SN events include GRB020903, GRB030329/SN2003dh, GRB031203/SN2003lw and GRB060218/SN2006aj from \citet{2008AJ....135.1136M}; GRB980425/SN1998bw \citep{2006A&A...454..103H}, GRB100316D/SN2010bh \citep{2011MNRAS.411.2792S}. }
\label{fig:L_Z_relation}
\end{figure*}

Fig.\,\ref{fig:L_Z_relation} shows the luminosity-metallicity relationship for nearby dwarf galaxies with measured abundances with the $T_{\rm e}$ method. We choose to plot only results from this method, given the 
large offsets and calibration issues when the strong lined methods are used e.g. the $R_{23}$ or N2 method
\citep{2011ApJ...729...56B}. 
There are only four SLSN Ic hosts which have oxygen abundances measured by the direct method: 
PTF12dam, SN~2010gx \citep{2013ApJ...763L..28C}, PS1-10bzj \citep{2013ApJ...771...97L}, and SN~2011ke \citep{2014ApJ...787..138L}. Since there is typically less than 0.1 mag difference between {\it B} and {\it g} magnitudes, we simply plot {\it M$_{g}$} on the same axis with no conversion for these four galaxies.
On this plot, the four SLSN hosts lie at the extreme lower end of the oxygen abundance distribution for their
equivalent {\it M$_{B}$}. 
In other words they appear not just to be low metallicity dwarf galaxies, but to have peculiarly low
oxygen abundances for dwarf galaxies. They occupy the same area as the four GRB host galaxies which have 
$T_{\rm e}$ determined oxygen abundances from \cite{2008AJ....135.1136M}. This quantitatively supports the results from the extensive sample in \citet{2014ApJ...787..138L} that SLSN Ic host galaxies are similar to those of GRBs. 
Although we should bear in mind the caveat that 
the four GRB hosts in Fig.\,\ref{fig:L_Z_relation} were selected by 
\cite{2008AJ....135.1136M} to be nearby and have associated supernovae detected. 
\citet{2014ApJ...787..138L} showed a similar plot of oxygen abundance vs host galaxy mass, which shows 15 
SLSN Ic hosts measured with the $R_{23}$ method. However due to the double value possibility in the $R_{23}$
method, many of the points have large uncertainties in their absolute values. 
The work here shows that when the $T_{\rm e}$ direct method is applied, these quite extreme low metallicities hold up to scrutiny. The conclusion of \citet{2014ApJ...787..138L} that they are quantitatively similar to GRB galaxies (at least low redshift GRBs) is supported by the comparison in Fig.\,\ref{fig:L_Z_relation} .

With an absolute magnitude of $M_{g} = -19.33\pm0.10$ the host of PTF12dam shows very similar properties to Green Pea galaxies, which are star-forming, luminous compact dwarf galaxies \citep{2009MNRAS.399.1191C}. It lies
close to the locus of points of Green Pea galaxies in Fig.\,\ref{fig:L_Z_relation} and the physical
parameters are similar. Green Peas typically have low mass($\sim 10^{8.8}$ \msun), high sSFR (1-100 Gyr$^{-1}$) and low-metallicity \citep[$\sim 0.2$ solar abundance; see ][]{2011ApJ...728..161I}. \citet{2010ApJ...715L.128A}, 
estimated a mean value of $12 + \log{\rm (O/H)} = 8.05 \pm 0.14$ for a sample of 79 star-forming Green Peas 
from the $T_{\rm e}$ method. The host of PTF12dam has the same metallicity as the mean value of Green Peas, a slightly lower stellar mass of $10^{8.4}$ \msun and a sSFR of 17 Gyr$^{-1}$ which is on the high end of the 
Green Pea distribution. 
In the \cite{1981PASP...93....5B} diagram, the host of PTF12dam is located in the starburst region with line ratios: $\log$(\Oiii$\lambda5007$/\hb) = 0.77 and $\log$ (\Nii$\lambda6583$/\ha) = $-1.41$. Again, this is similar to the Green Pea distribution and close to the starburst/AGN boundary area. Also Fig. 3 in \citet{2015MNRAS.449..917L} presented a comprehensive sample including SLSN, GRB hosts and EELGs, which supports this comparison.
The nitrogen/oxygen abundance ratio $\log{\rm (N/O)} = -1.38 \pm 0.13$ of the PTF12dam host is also similar to Green Peas \citep[see Fig.\,2 in][]{2010ApJ...715L.128A}. 
The high ionisation lines He\,{\sc II} $\lambda4686$ and \Neiii $\lambda3869$, are detected in the PTF12dam host galaxy which is unusual but not unprecedented in SN and GRB host galaxy spectra
\citep{2009ApJ...691..182S}.

\begin{figure}
\includegraphics[angle=0,width=\linewidth]{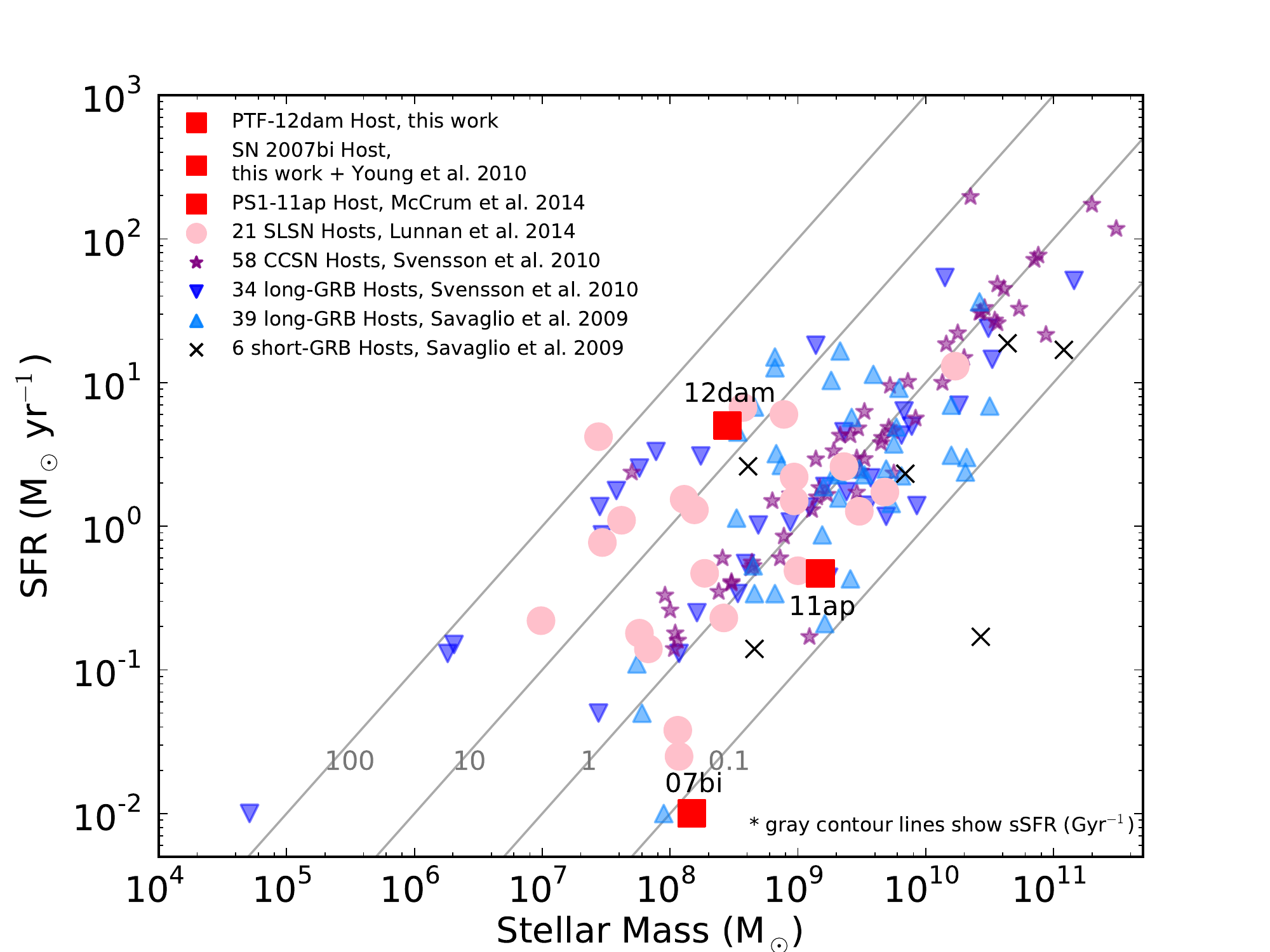}
\includegraphics[angle=0,width=\linewidth]{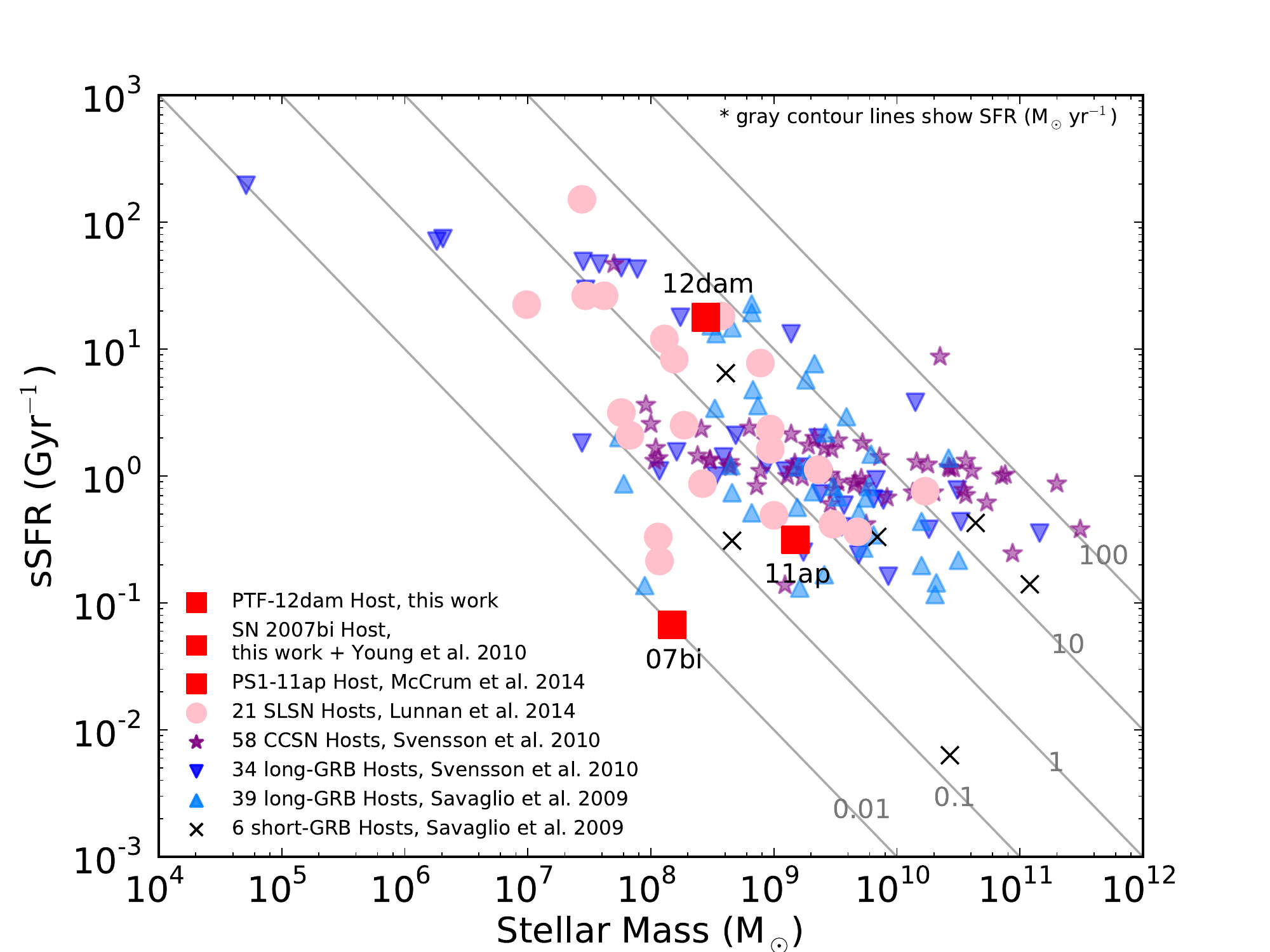}
\caption{Stellar mass vs. star-formation rate (top) and specific star-formation rate (bottom) for SLSNe Ic \citep{2014ApJ...787..138L}, GRB \citep{2009ApJ...691..182S, 2010MNRAS.405...57S} and CCSN hosts \citep{2010MNRAS.405...57S}. The three red boxes show the locations
of the three slowly fading SLSNe Ic published to date \citep{2010A&A...512A..70Y, 2014MNRAS.437..656M}.}
\label{fig:mass_SFR}
\end{figure}

The host of PTF12dam is a low mass galaxy of $2.8\times10^{8}$\,\msun, consistent with the mean mass value of 31 SLSN Ic hosts from \citet{2014ApJ...787..138L} of $2\times10^{8}$\,\msun. 
\citet{2014ApJ...787..138L} used the FAST stellar population synthesis code 
\citep{2009ApJ...700..221K} and fitted the SEDs with the stellar libraries of \cite{2005MNRAS.362..799M}. For PTF12dam they derived a value of $5.9^{+0.2}_{-0.5}\times10^{8}$\msun, which is a factor of two higher than what we estimate. In our detailed comparison of methods, we showed that although 
a wider wavelength coverage is useful, it is more critical to 
remove the contribution of emission lines to broad band 
photometric measurements when applying models 
based on stellar population synthesis (for example the {\sc MAGPHYS} code we employed here). Hence it is likely that the 
difference is due primarily to the inclusion of emission line flux in the broad band photometry used by 
\citet{2014ApJ...787..138L}, although our inclusion of UV and NIR data also improves the reliability of the mass inferred. 

By comparison, the average stellar mass of GRB hosts of 46 objects from \citet{2009ApJ...691..182S} is $2.0\times10^{9}$ \msun, and \citet{2010MNRAS.405...57S} derived a median mass from their 34 GRB host sample of $1.3\times10^{9}$ \msun derived from AB absolute {\it K}-band magnitude. Although it would initially appear that the host of PTF12dam is about 10 times less massive than these GRB host galaxies, 
the different methods employed can cause significantly different results. \citet{2009ApJ...691..182S} provided a calibration relation between stellar mass and {\it M$_{K}$} luminosity. Applying this, we would determine a mass for the PTF12dam host of $7.9-9.9\times10^{8}$ \msun (without and with {\it K}-correction for {\it M$_{K}$}), which is a factor of 2-3 more massive than that from the full SED fitting.
Hence comparison between results from different methods should be 
done with some caution. 

\citet{2014arXiv1411.1104T} also studied the host galaxy of PTF12dam. They used the OSIRIS spectroscopy at the 10.4-m GTC to observe the host after +567d from the SN peak. They claimed that the SN position was resolved in their long-slit image and found a very young ($\sim$ 3 Myr) stellar population at the SN site, which implied the progenitor of PTF12dam was a massive star ($>$ 60 \msun) from this recent starburst. However, there was no detection of the W-R features in their host spectra at the SN site. The W-R feature has been known to be a good tracer of recent massive star formation \citep{1981A&A...101L...5K}. There is a possible blue bump around 4640\AA\ in our late-time (+508d) spectrum of PTF12dam (see Fig.\,\ref{fig:12dam_GTC_spec}), which was taken from the GTC with 2.6 times longer exposure time (2400 sec $\times$ 3) than \citeauthor{2014arXiv1411.1104T}'s spectrum. These features are usually identified as {\ion{N}{iii}\,} and {\ion{C}{iii}\,} / {\ion{C}{iv}\,} emission lines from W-R stars. Nevertheless, our GTC spectrum was very noisy at this edge region, and we cannot confirm this feature. In addition, there are also no W-R bumps in our main WHT spectrum (Fig.\,\ref{fig:12dam_host_spec}). Although some \Feiii emission lines shown in the spectrum, they are very weak and narrow (e.g. EW of \Feiii $\lambda 4658 \sim 1.76$ \AA, \Feiii $\lambda 4986 \sim 1.36$ \AA). \citet{2008A&A...485..657B} analysed the galaxies with W-R features from the SDSS database, and found that the main distribution of the widths of W-R lines are from 1000 to 3000 km\,s$^{-1}$ FWHM from the \Heii $\lambda 4686$ line. We measured the width of the \Heii line in our WHT spectrum to be $\sim$ 300 km\,s$^{-1}$ FWHM. In summary there are no clear detections of the W-R features in our spectra, neither the broad W-R lines nor a broad component in the \Heii line.

\begin{table*}
\centering
\caption{Host galaxy properties comparison of the three 2007bi-like SNe known to date.}
\label{tab:07bi_like}
\begin{tabular}[c]{lllllllllll}
\hline
 SN & z &  M$_{\it g}$ & A$_{\it V}$ & D & SFR & M & sSFR & 12 + $\log$(O/H) ($T_{\rm e}$) & 12 + $\log$(O/H) (M91) & Reference \\
  & & (mag) & (mag) & (kpc) & (\msun\,yr$^{-1}$) & (\msun) & (Gyr$^{-1}$) & (dex) & (dex) &  \\
\hline
 PTF12dam & 0.107 &  $-19.33\pm0.10$ & 0.2 & $\sim$ 1.9 & 5.0 & $2.8\times10^{8}$ & 17.31 & $8.05\pm0.09$& $8.18\pm0.03$ & (1) \\
 SN~2007bi & 0.127 & $-16.05\pm0.10$ & 0 & $\sim$ 2.1 & 0.01 & $1.4\times10^{8}$ & 0.07 & - & 8.14 / 8.50 & (2) \\
 PS1-11ap & 0.524 &  $-17.79\pm0.32$ & 0 & $\sim$ 1.5 & $0.47\pm0.12$ & $1.5\times10^{9}$ & 0.31 & - & 8.1 / 8.5 & (3) \\
\hline 
\end{tabular}
(1) this work; (2) this work and \cite{2010A&A...512A..70Y} ; (3) \cite{2014MNRAS.437..656M}
\end{table*}

We investigated three separate methods to determine the SFR in the host of PTF12dam and show that the SFR from 
H$\alpha$ (5.0 \msun\,$yr^{-1}$) is different to that from the FUV (1.8 \msun\,$yr^{-1}$).
This might not necessarily be physically consistent. The H$\alpha$/H$\beta$ flux ratio could suggest that the photons arising in H\,{\sc ii} regions suffer from different extinction to continuum FUV photons as traced by the stellar continua of B-type stars.
We also showed that using \Oii luminosity in this type of star-forming galaxy 
gives a result that is consistent (5.2 \msun\,$yr^{-1}$) with that from \ha luminosity. This is useful when working at higher redshifts ($z > 0.5$) when H$\alpha$ is shifted to the NIR and \Oii $\lambda$3727 is still accessible in the optical. The large SFR for PTF12dam 
is one of the highest measured for a SLSN Ic hosts \citep{2014ApJ...787..138L}, which is not unexpected given the strength of the emission lines. When combined with the mass measurement, the PTF12dam host galaxy has a very high specific star-formation rate of 
sSFR = 17.9 Gyr$^{-1}$, which is higher than that of field galaxies at the same redshift. 
Fig.\,\ref{fig:mass_SFR} shows a comparison between GRB and CCSN hosts, SLSN Ic and the three 2007bi-like SN hosts that have been published to date. The host of PTF12dam has a high star-forming activity compared to the hosts of SN~2007bi and PS1-11ap. The stellar masses are quite similar although PS1-11ap is more massive. The sSFR of PTF12dam host is high, 50 times that of PS1-11ap's and up to 250 times higher than SN~2007bi's host.

The physical parameters of the hosts of these SN~2007bi-like SNe are also listed in Table\,\ref{tab:07bi_like}. 
We recalculated the physical size (D) of the PS1-11ap host galaxy of $\sim 1.5$ kpc using the angular size distance 1255 Mpc instead of the luminosity distance (2914 Mpc) used in \citet{2014MNRAS.437..656M}.
For SN~2007bi and PS1-11ap
we had to use the $R_{23}$ method since the \Oiii $\lambda$4363 line was not detected. We adopted the low branch solution since $\log$(\Nii/\Oii) $<$ 1.2 \citep[following][]{2002ApJS..142...35K}. The hosts of PTF12dam, PS1-11ap and SN~2007bi span the full range of stellar mass and SFR that other SLSNe Ic occupy. They are all star-forming dwarf galaxies, with low metallicity and young stellar populations but they don't appear to be significantly different (nor self-similar) in comparison to the 
other SLSNe Ic. One should recall that SN~2007bi was initially suggested to be a pair-instability explosion by \cite{2009Natur.462..624G} and PS1-11ap and PTF12dam are extremely similar in their properties. These three are the only well studied SLSNe Ic which have slowly fading lightcurves and are the only possible pair-instability SNe 
candidates in the local Universe with published data. However we can conclude that their host environments do not show any systematic difference from the bulk of SLSNe Ic.

The lowest metallicity host of any type of supernova to date has been found to be that of SN~2010gx 
at $12 + \log{\rm (O/H)} = 7.4 \pm 0.1$ ($ Z = 0.05 {\rm Z}_{\odot}$) on the $T_{\rm e}$ method scale \citep{2013ApJ...763L..28C}. The host of PTF12dam also follows this low metallicity trend. 
As discussed by \citet{2013ApJ...763L..28C} and \citet{2014ApJ...787..138L}, it appears that low
metallicity is a fundamental requirement to producing SLSNe Ic. Only one out of the \citeauthor{2014ApJ...787..138L} sample has a possible 
normal, solar like, metallicity (MLS121104) but this estimate is based on $R_{23}$ and the SN is significantly offset from the possible host. The agglomeration of current evidence is now strongly suggestive that 
low metallicity, with a critical value of around 8.0\,dex is required to produce SLSNe Ic, including the
fast declining SN~2005ap/SN~2010gx types and the more slowly evolving SN~2007bi/PTF12dam type explosions.

\section{Conclusions}
\label{sec:conclusions}

The host galaxies of SLSNe Ic seem to be exclusively compact, low mass, highly star-forming and of low metallicity. This poses a practical problem for detailed studies of both the host galaxy and late evolution of the supernova. Disentangling the flux of the SN and that of the host requires deep imaging over a period of several years to be sure that uncontaminated host flux is recovered and can be used as a reliable template to recover the faint and fading supernova. High spatial 
resolution imaging with HST, as done by 
\cite{2014ApJ...787..138L} can help resolve the ambiguity. 
In this paper we demonstrated that it can be done from ground based imaging, even in the case of a relatively bright, marginally resolved host galaxy with the SN occurring in the core. 
 We demonstrated recovery of the SN flux when it is around 1-2\% 
of the host galaxy light. 

The data presented here extend the lightcurve of PTF12dam out to +400d after peak allowing further testing of the three competing models for SLSNe Ic. The new data points indicate that 
PTF12dam declines faster than expected from 
the study of the first 200 days by \citet{2013Natur.502..346N}. 
We find that published pair-instability models cannot quantitatively fit full the bolometric lightcurve, 
as originally suggested in \citet{2013Natur.502..346N}.
The magnetar model is still viable, but the opacity in the ejecta needs
to be lowered to $\kappa_{\gamma} \simeq 0.01$ cm$^{2}$g$^{-1}$. 
This is physically plausible if the high energy radiation that provides the power is dominated by high energy gamma-rays produced via pair-production. 
This is an area that should be explored further with more detailed theoretical treatment. 
Finally, the model of interaction of the ejecta with a dense
CSM which thermalizes the kinetic energy and powers the
bright event fits the full bolometric lightcurve quite well. However as discussed in 
\cite{2014MNRAS.444.2096N}, the configuration of a 13 \msun CSM shell (which is H and He free) and 29 \msun of ejecta is hard to reconcile with plausible stellar progenitors. 
 Wherever the extra energy comes from to power these types of super-luminous supernovae, the progenitors are almost certainly carbon-oxygen stars in very metal poor galaxies. 

We derive a reliable value for the metallicity of the galaxy of 
$12 + \log({\rm O/H}) = 8.05 \pm 0.09$\,dex from the direct, electron temperature 
method (via detection of the weak [O\,{\sc iii}] $\lambda4363$ line). This is the fourth measurement of 
oxygen abundance in a SLSN host with this $T_{\rm e}$ method and all 
values are between 7.5-8.0\,dex. This supports the previous suggestions of 
\cite{2013ApJ...763L..28C} and \cite{2014ApJ...787..138L}
that low metallicity is a requirement to produce the 
stellar progenitor systems that give rise to these peculiar
SNe. Their rarity may be explained by this requirement to 
be formed at very low metallicity.

\section*{Acknowledgments}
The research leading to these results has received funding from the European Research Council under the European Union's Seventh Framework Programme (FP7/2007-2013)/ERC Grant agreement n$^{\rm o}$ [291222] (PI : S. J. Smartt). 
T.-W. Chen thanks to Daniel Kasen for their PISN models, and expresses appreciation to Yen-Chen Pan, Kai-Lung Sun, Meng-Chun Tsai, Chorng-Yuan Hwang, Max Pettini, Mark Sullivan, Gary J. Ferland, Edward Schlafly, and David R. Young for their useful advice. After the submission to the archive, this article has benefited greatly from the helpful feedback from the referee and Giorgos Leloudas, Avishay Gal-Yam, Alan Fitzsimmons, Michel Dennefeld and Paul Vreeswijk. 
R.P.K. and F.B. acknowledge support by the National Science Foundation under grant AST-100878798. N.E.R. acknowledges the support from the European Union Seventh Framework Programme (FP7/2007-2013) under grant agreement n. 267251 Astronomy Fellowships in Italy (AstroFIt). M.F. is supported by the European Union FP7 programme through ERC grant number 320360.

This work is based on observations collected at the William 
Herschel Telescope (WHT), operated on the island of La Palma 
by the Isaac Newton Group of Telescope; 
the Liverpool Telescope (LT), which is operated by 
LiverpoolJohn MooresUniversity in theSpan-ish Observatorio 
del Roque de los Muchachos of the Instituto de Astrofisica de 
Canarias with financial support from the UK Science and 
Technology Facilities Council; 
the Gran Telescopio Canarias (GTC), instaled in the Spanish 
Observatorio del Roque de los Muchachos of the Instituto de 
Astrofísica de Canarias, in the island of La Palma;
the Nordic Optical Telescope (NOT), operated by the Nordic Optical Telescope Scientific Association at the Observatorio del Roque de los Muchachos, La Palma, Spain, of the Instituto de Astrofisica de Canarias.



%

\label{lastpage}

\end{document}